\documentclass[%
reprint,
amsmath,amssymb,
aps
]{revtex4-2}

\usepackage{graphicx} 
\usepackage{subfigure}
\usepackage{array}
\usepackage{makecell}
\usepackage{comment}
\usepackage{dcolumn}
\usepackage{bm}
\usepackage{stmaryrd}
\usepackage{algpseudocode}
\usepackage{algorithm}
\usepackage{multirow}
\usepackage{acronym}
\usepackage{xcolor}
\usepackage{amsmath}
\usepackage{mathtools}

\begin{document}

\title{
Generalizing Egocentric Temporal Neighborhoods to probe for spatial correlations in temporal networks and infer their topology
}

\author{Didier Le Bail}
\thanks{Author to whom any correspondence should be addressed. Email address: didier.lebail@cpt.univ-mrs.fr}
\affiliation{Aix Marseille Univ, Université de Toulon, CNRS, CPT, Turing Center for Living Systems, Marseille, France}
\date{\today}

\begin{abstract}
Motifs are thought to be some fundamental components of social face-to-face interaction temporal networks.
However, the motifs previously considered are either limited to a handful of nodes and edges, or do not include triangles, which are thought to be of critical relevance to understand the dynamics of social systems.
Thus, we introduce a new class of motifs, that include these triangles, are not limited in their number of nodes or edges, and yet can be mined efficiently in any temporal network.
Referring to these motifs as the edge-centered motifs, we show analytically how they subsume the Egocentric Temporal Neighborhoods motifs of the literature.
We also confirm in empirical data that the edge-centered motifs bring relevant information with respect to the Egocentric motifs by using a
principle of maximum entropy.
Then, we show how mining for the edge-centered motifs in a network can be used to probe for spatial correlations in the underlying dynamics that have produced that network.
We deduce an approximate formula for the distribution of the edge-centered motifs in empirical networks of social face-to-face interactions.
In the last section of this paper, we explore how the statistics of the edge-centered motifs can be used to infer the complete topology of the network they were sampled from.
This leads to the needs of mathematical development, that we inaugurate here under the name of graph tiling theory.
\end{abstract}

\maketitle


\section{introduction}

We call motifs, or patterns, recurrent sub-elements that we find in data.
Counting how many motifs of each type occur in a sample of data often gives us insight about it.
Indeed, a system can often be described more simply as a combination of the motifs it contains.
For example, in complex systems, a motif can be associated with a precise low-level function.
Decomposing a system into its motifs allows us to decipher the various higher-level functions the system realizes.
For example, in the brain, different recurrent neural circuits are associated with various functions such as different forms of memory \cite{squire2009memory} or semantics \cite{pulvermuller2013semantic,amalric2018cortical} and many others.
In proteins, 3D sub-structures recurrent across all proteins are called functional motifs:
they perform each a specific biological function \cite{tyson2010functional,sobhy2016review,savojardo2023finding}.
Identifying these functional motifs inside a given protein allows us to predict its possible functions.
This key role of motifs as elementary building blocks of more complex systems motivates the development of methods to mine efficiently for them.
However, complex systems are so diverse that it could be thought that specific mining algorithms should be designed for each one of them.
This is not the case, because many complex systems can be modelled by the same object:
a graph, where nodes are parts of the system and edges are interactions between those parts.
For dynamical systems, interactions change over time, which results in varying ties in this graph representation.
In this case, the representation is called a temporal graph \cite{holme2012temporal,masuda2016guide,holme2015modern}, or temporal network, in opposition to a static network, whose nodes and edges are fixed in time.
In this unifying representation, the problem of finding motifs in many complex systems reduces to the problem of finding motifs in a temporal graph.
Many methods and algorithms were proposed to solve this problem both in the static and temporal setting \cite{paranjape2017motifs,liu2021temporal,kovanen2011temporal}, but a generic motifs mining algorithm has to face a computational obstacle:
motifs in a graph are sub-graphs.
Hence, counting motifs implies identifying all isomorphic sub-graphs.
As no algorithm is known to solve this problem in polynomial time, the motifs usually mined are restricted in their number of nodes or edges.
In the temporal setting, the number of possible motifs is much larger, which makes mining even more difficult.
Moreover, restricting to the sub-temporal graphs containing only a few edges does not give us access to the building blocks of complex systems, in particular of social face-to-face interaction temporal networks.
Indeed, the presence of burstiness in social interactions \cite{Moinet_2015,stehle2010dynamical} involves that an individual may interact with an arbitrarily large number of partners in a short time window.
That is why the motifs to look for in these temporal networks should not be limited in their number of nodes or edges.
Such tractable motifs have been found \cite{Longa2022}:
they are made of the interactions an ego has with its alters for a small number of time steps.
These motifs are originally called Egocentric Temporal Neighborhood (ETN).
Because they are limited in their duration and because they do not include the interactions between the alters, they can be mined for in linear time and size of the temporal network.
The ETN have been shown to be relevant building blocks of social face-to-face interaction networks \cite{longa2024generating}, probably because they give the ego's point of view, which reflects the social behaviour of the individual.
However, they are still not enough to understand social systems.
Indeed, they do not give any access to the groups of discussions and higher-order structures, which are omnipresent in social systems \cite{benson2016higher}.
That's why we define in this paper a new class of motifs together with an efficient mining algorithm, collecting both the ego's point of view and the interactions between its alters.
This allows us to extract motifs that contain triangles, which are the smallest possible non-trivial higher-order structures.

This paper is organized as follows:
In a first part, we describe in detail the ETN motifs of \cite{Longa2022}, that we rename as node-centered motifs (NCTN, for Node-Centered Temporal Neighborhood).
We then introduce a new type of motifs, called the edge-centered motifs, that include triangles.
We abbreviate these motifs as ECTN, for Edge-Centered Temporal Neighborhood.
In a second part, we investigate analytically how the ECTN and the NCTN motifs are related to each other.
We derive a formula that expresses the NCTN motifs distribution in terms of the ECTN motifs distribution, before using a principle of maximum entropy to infer the reverse:
the ECTN motifs distribution in terms of the NCTN motifs distribution.
In a third part, we compute analytically the ECTN distribution under various assumptions on the dynamics generating a temporal network.
We also explain how to compare our predictions with an ECTN distribution sampled from a given temporal network.
The fourth part is devoted to the confrontation between theory and experiments:
first we show that the principle of maximum entropy fails in empirical networks, proving that ECTN motifs bring additional information with respect to the NCTN motifs in those networks.
Second, we investigate which assumption on the dynamics generating a temporal network produces an ECTN distribution similar to the distributions observed in empirical networks of social face-to-face interactions.
From this investigation, we can estimate how complex must be a generative model to reproduce the empirical distributions of motifs.
In a fifth and last part, we inaugurate a mathematical theory, that if developed could be used to partially infer the topology of a network from its motifs distribution.
We then conclude on possible future research about motifs.

Before we go on to the next section, let us explain briefly the conventions of language we will be using in this paper:
\begin{itemize}
    \item A static network, or static graph, can be viewed as a cloud of points, with some pairs of points being linked together.
    The points are called vertices and the links are called edges.
    \item A temporal network, or temporal graph, can be viewed as a succession in time of static graphs.
    Said otherwise, it is the data of a set of vertices with a set of edges that are either active or inactive across time.
    \item In this paper, time is discrete and takes the form of an integer.
    \item If not specified, a graph is actually a temporal graph.
\end{itemize}

\section{definition of node and edge-centered motifs}
Before introducing edge-centered motifs, we begin by describing the motifs known under the acronym ETN, which stands for Egocentric Temporal Neighborhood.

\subsection{Egocentric Temporal Neighborhood}
ETN instances are sub-temporal graphs resulting from getting a node's point of view for a fixed number of consecutive time steps.
By ``node's point of view'', we mean the partners with whom the node interacts.

More precisely, an ETN instance is extracted as follows:\\
(1) we choose a node $i$ that we will focus on; this node is called the \textit{central node}\\
(2) we choose a duration $d$ of observation called the \textit{depth}; this is the number of consecutive time steps we will looking at the central node\\
(3) we choose a starting time $t$ for our observation\\
(4) we collect every interaction between the central node and any other node that occurs between the times $t$ and $t+d-1$\\

\begin{figure}
    \subfigure[temporal network]{
    \includegraphics[width=\columnwidth]{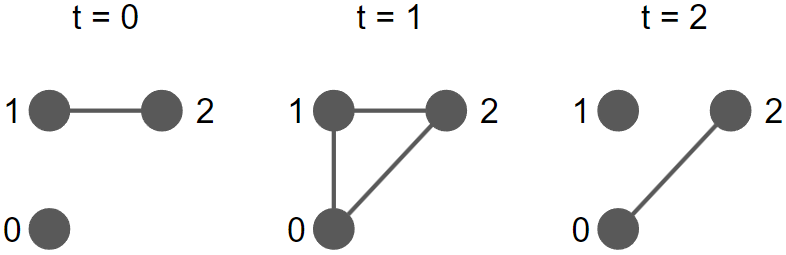}
    }
    \subfigure[ETN instances]{
    \includegraphics[width=\columnwidth]{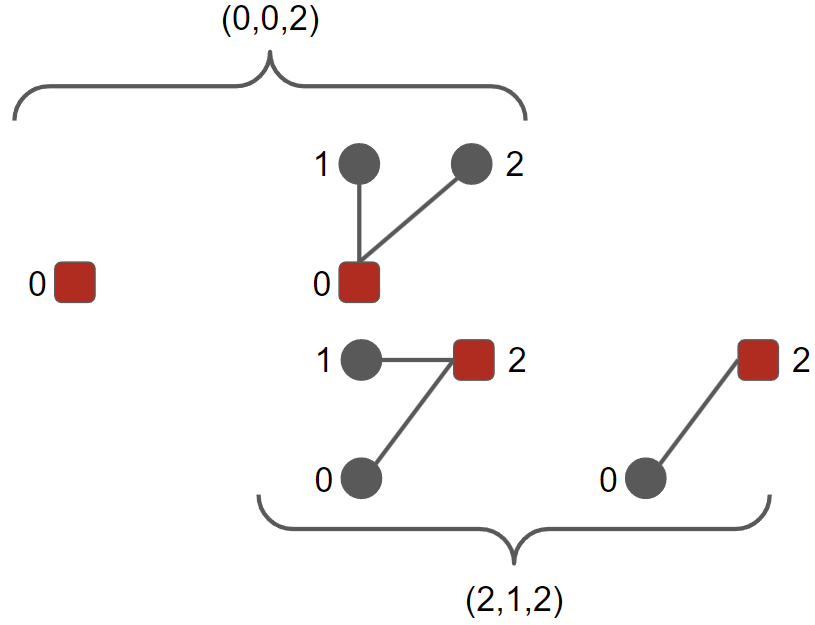}
    }
    \caption{\label{fig:0}\textbf{Examples of ETN instances.}
    (a) temporal network from which the ETN instances are extracted
    (b) two ETN instances.
    Each instance is a sub-temporal graph of the network in (a), labelled by the triple (central node, starting time, depth).
    Here the depth is two, meaning the duration of the sub-temporal graphs is equal to two.
    To facilitate reading, the central node has been represented as a red square instead of a gray disk.
    Note that the triangle at $t=1$ cannot be part of any ETN instance, as only the interactions between an ego (the central node) and its alters (the satellites) are collected.
    }
\end{figure}

Thus, an ETN instance is labelled by three variables:
the central node $i$, the starting time of observation $t$ and the depth $d$.
In practice, $d$ is usually taken to be 2 or 3 (see Figure \ref{fig:0} for an example of ETN instance).

Now we know what an ETN instance is, we will introduce the corresponding motif, so that we would be able to count how many instances there are of the same ETN motif in a graph.
Said otherwise, given a graph $G=(V,E,T)$ and the set of all ETN motifs $\mathcal{M}$, we want to compute the motif map:
$$
\begin{array}{cccc}
   m_{G}:  & V\times
T\times\mathbf{N} & \xrightarrow{} & \mathcal{M} \\
   & (i,t,d) & \mapsto & m_{G}(i,t,d)
\end{array}
$$
where $i,t,d$ are respectively the central node, starting time and depth of the instance of the ETN motif $m_{G}(i,t,d)$.

There are multiple ways of representing an ETN motif; we will describe three of them to help the reader getting familiar with this notion.

\subsubsection{Diagram representation}
\begin{figure}
    \includegraphics[width=\columnwidth]{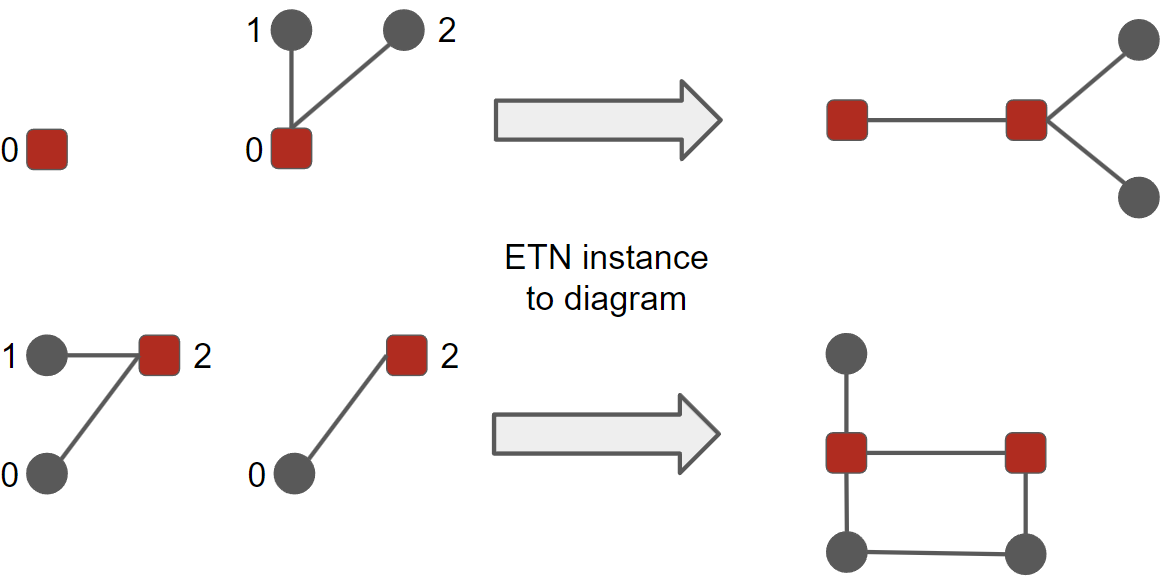}
    \caption{\label{fig:1}\textbf{Diagrammatic representation of ETN motifs.}
    When we go from the instance to the motif, the identity of the nodes is lost, as well as the starting time.
    The diagram representation is time-ordered:
    from left to right, the central node is drawn together with its satellites according to time ordering.
    An horizontal edge is drawn between two identical nodes at different time steps.
    A vertical edge is drawn between a satellite and the central node if they interact.
    }
\end{figure}

The most intuitive way of representing an ETN is to draw them as diagrams.
Such a diagram is a static graph with additional structure.
It has two types of nodes and two types of edges.
The nodes can either correspond to the central node or not, in which case they are called \textit{satellites}.
In an instance, satellites will correspond to the partners of the central node.
The edges can be either horizontal or vertical.
In an instance, horizontal edges are those drawn between two identical nodes at different time steps and vertical edges are those drawn between a satellite and the central node.
See Figure \ref{fig:1} for a diagrammatic representation of the motifs associated to the instances drawn in Figure \ref{fig:0}.

The diagrammatic representation is not the most compact representation of an ETN.
For example, the horizontal edges between the copies of the central node are redundant as they depend just on the depth.
Moreover vertical edges are only possible between a satellite and the central node.
From this, we see that the central node is not needed to represent an ETN.
This is illustrated in the next representation.

\subsubsection{String representation}
The reason why it is possible to mine for ETN efficiently is that they can be represented as binary strings.
Given an ETN, this string is built as follows:\\
(1) we build the \textit{activity profile} of each satellite.
In an ETN, the activity profile of a satellite is a string of length equal to the depth of the ETN made up with `0' and `1'.
In an instance $(i,t,d)$, a `1' at position $k$ means that the satellite is interacting at time $t+k$ with the central node $i$.\\
(2) we sort these activity profiles by lexicographic order (or any total order on the space of binary strings of length $d$)\\
(3) we concatenate these ordered profiles in a single string.

\begin{figure}
    \includegraphics[width=\columnwidth]{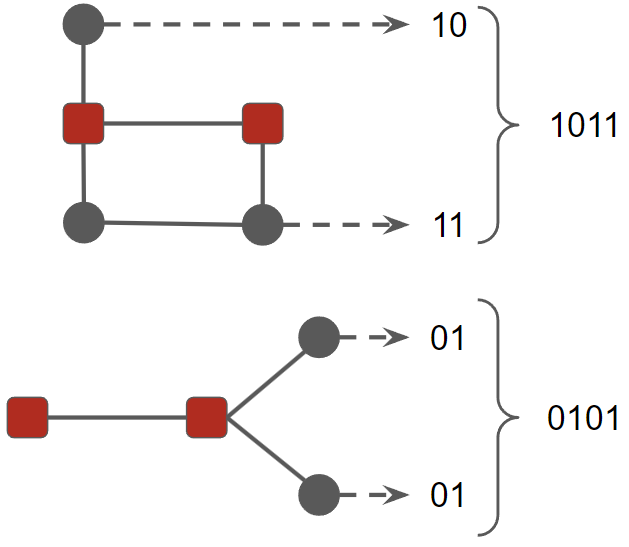}
    \caption{\label{fig:2}\textbf{Representation of an ETN motif into a binary string.}
    First the activity profile of each satellite is determined:
    a `1' means the satellite interacts with the central node.
    Then the profiles are sorted by increasing lexicographic order and concatenated into a single string of length $n_{s}d$, where $n_{s}$ is the number of satellites and $d$ is the ETN depth.
    }
\end{figure}

Note that any activity profile in an ETN contains at least one `1'.
Examples of this procedure are given in Figure \ref{fig:2}.

Note that the length of strings is variable and is \textit{a priori} unbounded for any depth greater than zero.
A representation of finite dimension is yet possible, as we will see just now.

\subsubsection{Map representation\label{subsubsec:map_rep}}
Note that contrary to the number of satellites, the number of possible activity profiles depends only on the depth.
Then, a representation of an ETN as a tuple with a finite number of components is possible as follows:
we count how many satellites realize each possible activity profile.
Then, denoting the set of possible activity profiles at depth $d$ as $A(d)$, an ETN of depth $d$ is equivalent to the map
$$(n_{a})_{a\in A(d)}$$
where $n_{a}$ is the number of satellites with $a$ as activity profile.

This map can equivalently be represented as a tuple as long as a total ordering convention is chosen on $A(d)$.
Note that the number of elements in $A(d)$ is:
$$|A(d)|=2^{d}-1$$
since the empty profile is not authorized.

\subsection{Edge-Centered Temporal Neighborhood}
We have defined what an Egocentric Temporal Neighborhood (ETN) is.
In the following, we will call equivalently these motifs as Node-Centered Temporal Neighborhood (NCTN), as opposed to the Edge-Centered Temporal Neighborhood (ECTN), that we are introducing now.

\begin{figure}
    \includegraphics[width=\columnwidth]{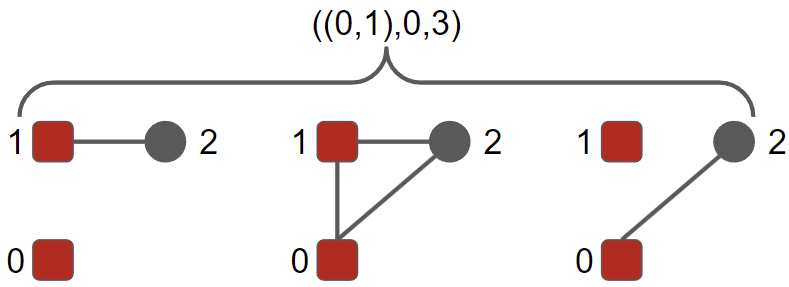}
    \caption{\label{fig:3}\textbf{Example of ECTN instance.}
    The ECTN instance is extracted from the same network as Figure \ref{fig:0}.
    Here the instance considered has (0,1) as central edge, starts at time 0 and its depth equals to 3.
    Note then that the whole temporal network is contained in the instance.
    }
\end{figure}

Extracting an instance of an ECTN motif amounts to focus on a pair of nodes, called the \textit{central edge}, and to collect every interaction between each node of the pair and any other node.
Like in the NCTN case, we want to know whether a node is the same from one time step to the other.
To be able to extract triangles, we also want to know, at each time step, whether the two nodes of the central edge are interacting with the same partner.
Like NCTN instances, an ECTN instance is labelled by a triple of the form (central object, starting time, depth).
For an ECTN instance, this triple is denoted by $(e,t,d)$, where $e\in E$ is the central edge, $t\in T$ is the starting time and $d\geq1$ the depth.
On Figure \ref{fig:3}, we show an example of ECTN instance.

\subsubsection{Diagram representation}

\begin{figure}
    \includegraphics[width=\columnwidth]{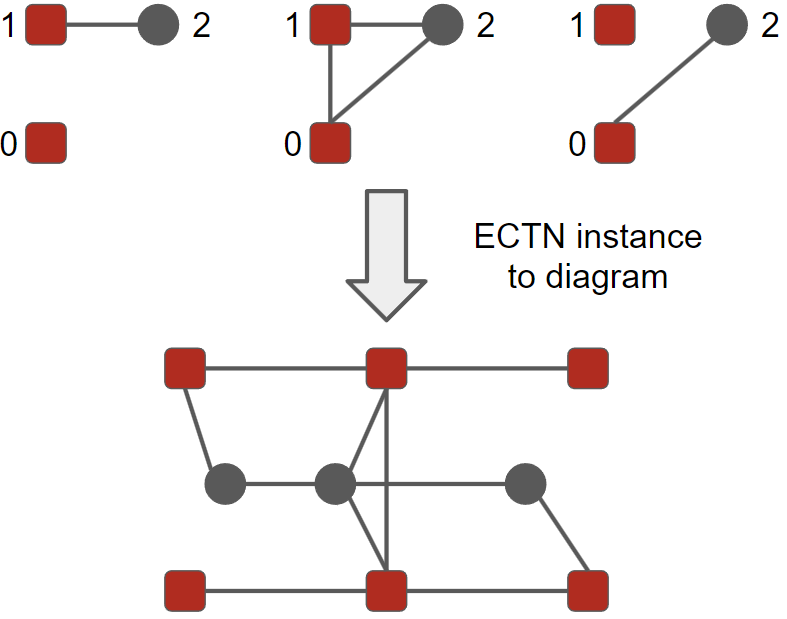}
    \caption{\label{fig:4}\textbf{Diagram representation of an ECTN motif.}
    We consider the same instance as in Figure \ref{fig:3}.
    From this instance, we build the diagram representing the underlying motif.
    Like in NCTN, horizontal edges are drawn between identical nodes at different time steps, and vertical edges indicate interactions between nodes.
    }
\end{figure}

Like in the NCTN case, we are interested in counting the number of instances of each ECTN motif in a given temporal network.
This motif can be represented as a diagram, by following the same rules of the drawing of a NCTN motif diagram (see Figure \ref{fig:4} for the drawing of an ECTN motif diagram).

Note that, contrary to the NCTN case, the central edge is no more redundant in the diagram representation, because even in the absence of satellites, it can be either active or inactive.
Said otherwise, the activity profile of the central edge cannot be deduced from the profiles of its satellites.
Despite this higher complexity, it is still possible to write such diagrams as strings in a fast way.

\subsubsection{string representation}
In the NCTN case, the central object needed not to be encoded.
Moreover, at each time step, a satellite could be either active, meaning the satellite is interacting with the central node at that time, or inactive.

In the ECTN case, the central object is encoded by an activity profile as defined previously:
a binary string of length $d$ with the `1' corresponding to the times of activity of the central edge.
At each time step, satellites can now be in either of four states.
The letter used to code each state is indicated in parentheses:
\begin{itemize}
    \item inactive (`0'): the satellite is not interacting with any node of the central edge
    \item active1  (`1'): the satellite is interacting with the first node of the central edge and not with the second
    \item active2  (`2'): the satellite is interacting with the second node of the central edge and not with the first
    \item active3  (`3'): the satellite is interacting with both nodes of the central edge
\end{itemize}

To distinguish between the second and third states, we need to distinguish between the two nodes of the central edge.
If we do not, we cannot recover the correct ECTN diagram from the string representation, and we would have multiple possible strings for a single ECTN motif.
We will see later how to solve this problem.

Once the profiles of the central edge and each satellite have been computed, we sort the satellite profiles by lexicographic order just like in the NCTN case, then we concatenate them into a single string.
Then we add the profile of the central edge at the left of this string.

For example, let us denote by $s$ the number of satellites, $a_{c}$ the profile of the central edge and $a_{1},\ldots,a_{s}$ the profiles of the satellites.
To obtain the ECTN string, we sort the $a_{i}$ by increasing order, i.e. we find a permutation $\pi$ such that
$$\forall 1\leq i\leq s-1, a_{\pi(i)}\leq a_{\pi(i+1)}$$
The final string reads
$$a_{c}a_{\pi(1)}\cdots a_{\pi(s)}$$

We are now ready to solve the attribution problem evoked earlier:
how to decide which node of the central should be associated the letter `1' or `2'?
All we need is to test both hypotheses and retain the one yielding the smallest final string.
Said otherwise, before sorting the activity profiles, we build another set of profiles where the letters `1' and `2' have been exchanged.
Then we compare the strings obtained after sorting and concatenation of these two sets of profiles, and return the smallest string with respect to the lexicographic order.

\begin{figure}
    \centering
    \includegraphics[width=\columnwidth]{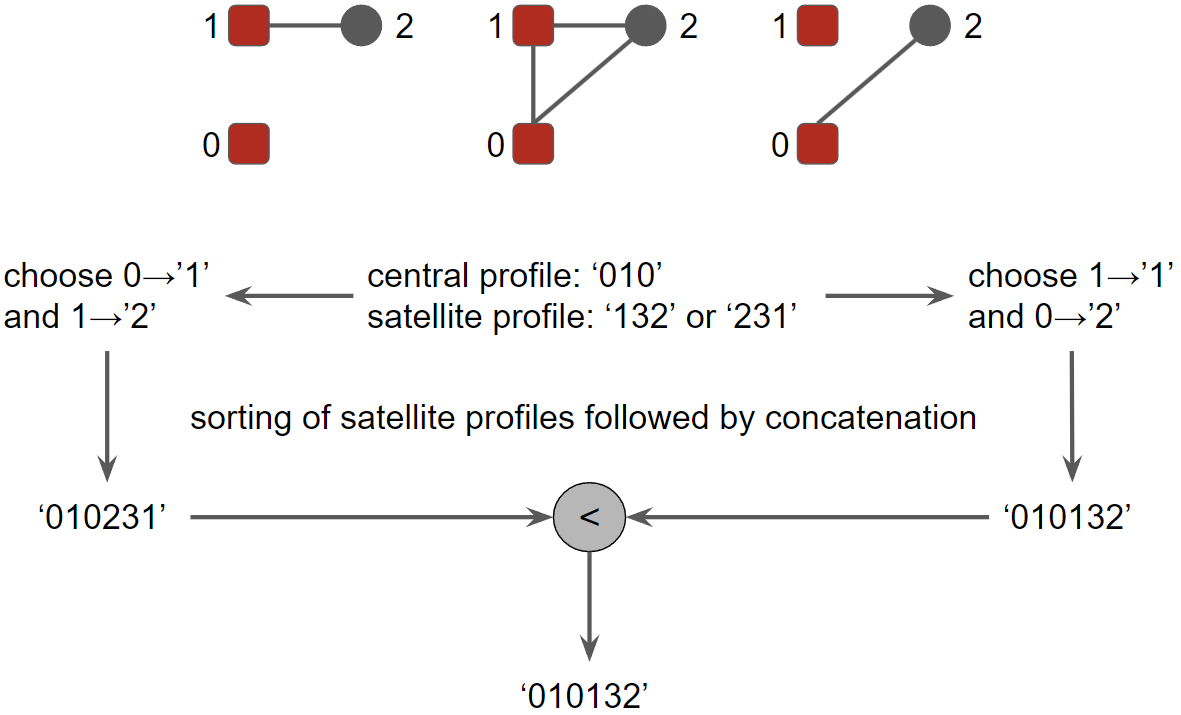}
    \caption{\label{fig:string_rep_ECTN}\textbf{String representation of an extracted ECTN.}
    The ECTN instance we take as an example is the same as in Figure \ref{fig:3}.
    At first, we do not know which naming convention to use.
    The first possibility is to encode by the letter `1' an exclusive interaction with the node 0, and by the letter `2' an exclusive interaction with the node 1.
    The second possibility is the reverse, obtained from the first by exchanging the letters `1' and `2' in the profiles of each satellite.
    We take as final string the smallest between the two possible ones.
    }
\end{figure}

We illustrate the conversion of an ECTN instance into its string on Figure \ref{fig:string_rep_ECTN}.

\section{relation between node and edge centered motifs}

What information do our motifs convey about a temporal network?
In particular, we will focus on the motifs distribution, i.e. the knowledge for every motif, of its number of instances in a temporal network.
What does this knowledge teach us about a temporal network?

\subsection{From ECTN to NCTN}
Before answering that question, we will show that the distribution of NCTN can be deduced from the distribution of ECTN, proving that the latter contain \textit{a priori} more information than the former.
Given a temporal network, let us denote by $N_{n}(m)$ the number of instances of the NCTN $m$ of depth $d$, and by $N_{e}(M)$ the number of instances of the ECTN $M$ of depth $d$.
We denote also by $m(i,t)$ the NCTN motif of instance $(i,t,d)$, and by $M(ij,t)$ the ECTN motif of instance $((i,j),t,d)$.
We have
$$N_{n}(m)=\sum_{i\in V,t\in T}\delta_{m,m(i,t)}$$
Introducing Sat($i,t$) the set of the satellites of the NCTN instance $(i,t,d)$, we have:
$$
N_{n}(m)=\sum_{i\in V,t\in T}\delta_{m,m(i,t)}\frac{d}{|m|}\sum_{j\in\text{Sat($i,t$)}}\sum_{M}\delta_{M,M(ij,t)}
$$
Indeed, the sum $\sum_{M}$ denotes the sum over all ECTN motifs and $|m|$ is the length of the NCTN $m$ viewed as a string.
Then the ratio $\frac{|m|}{d}$ is nothing but the number of satellites of $m$.
Reorganizing the terms, we get:
$$N_{n}(m)=\frac{d}{|m|}\sum_{i\in V,t\in T}\sum_{j\in\text{Sat($i,t$)}}\sum_{M}\delta_{M,M(ij,t)}\delta_{m,m(i,t)}$$
Now, iterating over the indices $i,t,j$ of the first two sums amounts to this:
\begin{enumerate}
    \item pick a node $i$ and a time $t$
    \item pick a node $j$ such that the edge $(i,j)$ is active at least once between the times $t$ and $t+d-1$ included.
\end{enumerate}
This iteration is equivalent to an iteration over instances $(i,j,t)$ of non-empty ECTN motifs of depth $d$, on condition that we count twice each instance.
Denoting by $I_{d}^{M}$ the set of such instances, we get:
$$N_{n}(m)=2\frac{d}{|m|}\sum_{(ij,t)\in I_{d}^{M}}\sum_{M}\delta_{M,M(ij,t)}\delta_{m,m(i,t)}$$
where $ij$ denotes the undirected edge $(i,j)$.
This can be symmetrized as:
$$N_{n}(m)=\frac{d}{|m|}\sum_{(ij,t)\in I_{d}^{M}}\sum_{M}\delta_{M,M(ij,t)}\left(\delta_{m,m(i,t)}+\delta_{m,m(j,t)}\right)$$
The term $\delta_{m,m(i,t)}$ can be rewritten as a condition on $M$:
if $i$ and $j$ are the nodes of the central edge of $M$, either the NCTN centered on $i$ or the NCTN centered on $j$ should be equal to $m$.

We can extract a NCTN motif from a ECTN motif by introducing maps $\psi_{1}$ and $\psi_{2}$ defined as follows:
$\psi_{1}$ (resp. $\psi_{2}$) takes an ECTN motif and returns the NCTN motif centered on node of the central edge associated to the letter `1' (resp. `2').
In practice, this can be done by using the maps $\phi_{1}$ and $\phi_{2}$ defined later.
Back to our formula, we obtain:
$$N_{n}(m)=\frac{d}{|m|}\sum_{k=1}^{2}\sum_{M\in\psi_{k}^{-1}(m)}\sum_{(ij,t)\in I_{d}^{M}}\delta_{M,M(ij,t)}$$
This is nothing less than:
$$N_{n}(m)=\frac{d}{|m|}\sum_{k=1}^{2}\sum_{M\in\psi_{k}^{-1}(m)}N_{e}(M)$$
This can be written in a more compact way if we use the notion of disjoint union $\coprod$.
A disjoint union of sets is their union taking into account the multiplicity of elements.
For example, $\{x\}\coprod\{x\}$ contains two elements.
Thus we have:
\begin{equation}\label{eq:ectn_to_nctn}
    N_{n}(m)=\frac{d}{|m|}\sum_{M\in E_{m}}N_{e}(M)
\end{equation}
where $E_{m}=\psi_{1}^{-1}(m)\coprod\psi_{2}^{-1}(m)$.

Now, what if we are interested in the motifs frequencies and not in their number of instances?
The relation between the two is the partition function:
$$
\begin{cases}
    Z_{n}=\sum_{m\in\{\text{NCTN}\}}N_{n}(m)
    \\
    Z_{e}=\sum_{M\in\{\text{ECTN}\}}N_{e}(M)
\end{cases}
$$
From that we deduce
$$P_{n}(m)=\frac{dZ_{e}}{|m|Z_{n}}P_{e}(M\in E_{m})$$

Is there a formula relating the two partition functions that does not require the knowledge of $N_{e}$?

We will show that $Z_{n}$ and $Z_{e}$ are related to the node and edge interduration distributions, respectively.
As these distributions are decoupled from each other in the general case, this might indicate that there is no simple expression of $Z_{n}$ in terms of $Z_{e}$.

Let us first consider the NCTN partition function:
$Z_{n}$ is the number of couples $(i,t)$ such that the node $i$ is active at least once within the window of size $d$ starting from time $t$.
Then, if we add to $Z_{n}$ the number of couples $(i,t)$ that are never active within the same window, we get all possible couples.

\begin{figure}
    \centering
    \includegraphics[width=\columnwidth]{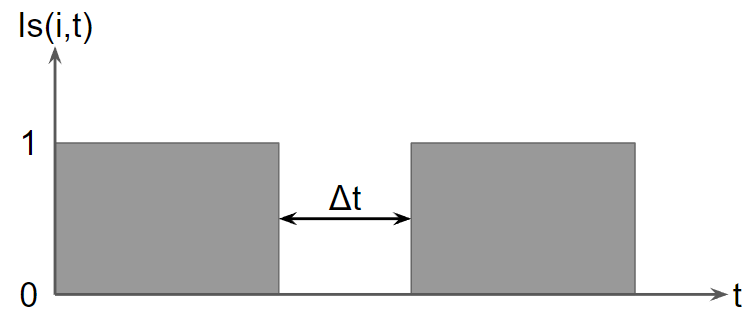}
    \caption{\label{fig:Z_NCTN}\textbf{Computation of the NCTN partition function.}
    We plot an example of activation function of a node $i$ versus time.
    The function returns 1 at $t$ if $i$ is active at $t$ and 0 else.
    The average size of the gap between two consecutive periods of activity, denoted by $\Delta t$ on the plot, is related to the NCTN partition function $Z_{n}$ (see text). 
    }
\end{figure}

The difference $|V||T|-Z_{n}$ can be expressed in terms of the average number of consecutive times a node is inactive:
Consider a node $i$ and let us plot its activity Is($i,t$) over time on Figure \ref{fig:Z_NCTN}), where Is($i,t$) returns True if the node $i$ is active at time $t$, and False else.

On this plot, we introduce the gap between two consecutive periods of activity of $i$, that we write $\Delta t$.
The length of such periods of inactivity is called node interduration.
Since $\Delta t$ is the number of times of inactivity, for each interduration period of size $\Delta t\geq d$, there are $\Delta t-d+1$ times $t$ within this period, such that $(i,t)$ is inactive for at least $d$ consecutive time steps.

Formally, let us denote by $N_{ni}(\tau)$ the number of periods of node inactivity of size $\tau$.
We have:
$$|V||T|-Z_{n}=\sum_{\tau\geq1}(\tau-d+1)\theta(\tau-d+1)N_{ni}(\tau)$$
Thus, introducing the partition function of the node interduration distribution, we deduce that
$$|V||T|-Z_{n}=Z_{ni}\left<(\tau-d+1)\theta(\tau-d+1)\right>_{P_{ni}(\tau)}$$
where $\left<g(x)\right>_{P(x)}$ denotes the average of $g$ with respect to the distribution $P$ for the random variable $x$.

By a similar computation, we can show that the ECTN partition function satisfies:
$$\left|\bigcup_{t\in T}E_{t}\right||T|-Z_{e}=Z_{ei}\left<(\tau-d+1)\theta(\tau-d+1)\right>_{P_{ei}(\tau)}$$
where $P_{ei}$ denotes the edge interduration distribution and $Z_{ei}$ its partition function.
$E_{t}$ denotes the set of active edges at time $t$, so the union of those sets for every $t$ is the set of edges of the fully aggregated network.

From these two formula, we can conclude that the NCTN and ECTN partition functions cannot be deduced from one another.
Computing the NCTN partition function from the ECTN motifs would require the knowledge of $N_{e}(M)$ for each ECTN $M$.

\subsection{From NCTN to ECTN}
In the previous subsection, we have shown that the NCTN distribution could be deduced from the ECTN distribution.
However, this does not mean that the ECTN contain systematically more information than the NCTN.
It depends on the dynamics generating the network from which ECTN and NCTN are sampled:
in some temporal networks, all the relevant information are already grasped by NCTN and the ECTN appear to be random gluing of NCTN.
In such cases, we could hope to recover the ECTN distribution from the NCTN distribution.

In the general case, let us make it clear that this is not possible.
For example, it is possible to build an ECTN distribution compatible with a NCTN distribution via the following procedure:
\begin{enumerate}
    \item take a NCTN
    \item take one satellite
    \item consider the ECTN centered on the central edge given by the pair of this satellite and the central node of your NCTN, and that contains all the interactions visible in this NCTN
\end{enumerate}
For example, the NCTN of depth 2 and string `1011' will yield the two ECTN `1011' and `1110'.

While this construction is always possible, note that the resulting ECTN will never contain any triangle.
Indeed, their string representation will contain only `0' and `1', the letters `2' and `3' will be missing.

As a consequence, this ECTN distribution has little chance to match the actual ECTN distribution of the data set considered.

This being said, how to test whether ECTN bring new information with respect to NCTN for a particular data set?
If the distribution of ECTN does not bring any information, it should be as random as possible.
That is why we propose to compare the ECTN distribution sampled from a temporal network with the ECTN distribution of maximum entropy, constrained to be compatible with the NCTN distribution of that network.
If the two are similar, we will say that the ECTN distribution of that network satisfies the principle of maximum entropy (MEP).
It not, then information other than the NCTN distribution are needed to recover the ECTN distribution of that network.

Let us now compute the ECTN distribution with maximum entropy (the MEP distribution).

\subsubsection{Computation of the MEP distribution}
To do this, we use Lagrange multipliers to enforce the constraints under which the entropy should be maximized:
\begin{align*}
L(N_{e},\alpha,\beta)
&=
-\sum_{M}\frac{N_{e}(M)}{Z_{e}}\log\left(\frac{N_{e}(M)}{Z_{e}}\right)
\\
&+\alpha\left(\sum_{M}N_{e}(M)-Z_{e}\right)
\\
&+\sum_{m}\beta_{m}\left(N_{n}(m)-\frac{d}{|m|}\sum_{M\in E_{m}}N_{e}(M)\right)    
\end{align*}

Deriving with respect to $N_{e}(M)$ for each ECTN $M$ yields
\begin{align*}
\frac{\delta L}{\delta N_{e}(M)}
&=
-\frac{1}{Z_{e}}\left(1+\log\left(\frac{N_{e}(M)}{Z_{e}}\right)\right)
\\
&+\alpha
\\
&-\sum_{m}\beta_{m}\frac{d}{|m|}\sum_{M'\in E_{m}}\delta_{M,M'}
\end{align*}

Equating to zero gives the expression for $N_{e}$ in terms of the Lagrange multipliers:
\begin{multline}\label{eq:lag_zero}
\log\left(\frac{N_{e}(M)}{Z_{e}}\right)
=
Z_{e}\alpha-1-dZ_{e}\sum_{k=1}^{2}\frac{\beta_{\psi_{k}(M)}}{|\psi_{k}(M)|}
\end{multline}

What remains to do is obtaining the expression for $\beta_{m}$.
It can be found by injecting the expression for $N_{e}$ into the constraint on $N_{n}$.
Let us introduce
\begin{equation*}
\begin{cases}
    \gamma_{m}=\exp\left(-\beta_{m}\frac{dZ_{e}}{|m|}\right)
    \\
    \lambda=dZ_{e}\exp\left(Z_{e}\alpha-1\right)
\end{cases}
\end{equation*}

Then eq \ref{eq:lag_zero} rewrites as
\begin{equation*}
N_{e}(M)
=
\frac{\lambda}{d}\gamma_{\psi_{1}(M)}\gamma_{\psi_{2}(M)}    
\end{equation*}

The unknowns are the $\lambda$ and the $\gamma_{m}$, but we can reduce their number by introducing
$$
\tilde{\gamma}_{m}=\gamma_{m}\sqrt{\lambda}
$$

This leads to the more convenient form for eq \ref{eq:lag_zero}:
\begin{equation}\label{eq:ectn_lag}
N_{e}(M)
=
\frac{1}{d}\tilde{\gamma}_{\psi_{1}(M)}\tilde{\gamma}_{\psi_{2}(M)}    
\end{equation}
 
By injecting this expression into eq \ref{eq:ectn_to_nctn}, we get:
\begin{equation*}
\sum_{M\in E_{m}}\tilde{\gamma}_{\psi_{1}(M)}\tilde{\gamma}_{\psi_{2}(M)}
=
|m|N_{n}(m)
\end{equation*}

Let us introduce
$$
E_{m,m'}=\psi_{1}^{-1}(m)\cap\psi_{2}^{-1}(m')
$$

Then we can show that
\begin{equation*}
\tilde{\gamma}_{m}\sum_{m'}\left(|E_{m,m'}|+|E_{m',m}|\right)\tilde{\gamma}_{m'}
=
|m|N_{n}(m)
\end{equation*}

This system of equations can be written more conveniently if we introduce
$$
\begin{cases}
    \left(E_{S}\right)_{m,m'}=|E_{m,m'}|+|E_{m',m}|
    \\
    V_{m}=N_{n}(m)|m|
\end{cases}
$$

Then we get:
\begin{equation}\label{eq:lag_to_solve}
\tilde{\gamma}\cdot \left(E_{S}\tilde{\gamma}\right)
=
V
\end{equation}
where the dot $\cdot$ denotes the element-wise product between two matrices of same size.

Once the $\tilde{\gamma}_{m}$ have been computed, the ECTN distribution is deduced from eq
\ref{eq:ectn_lag}.
To solve the system of equations \ref{eq:lag_to_solve}, we first computed the matrix $E_{S}$ and then used the numerical solver ``scipy.optimize.least\_squares''\cite{byrd1988approximate,branch1999subspace,more2006levenberg} in an algorithm written in Python.

\subsubsection{The gluing matrices\label{subsubsec:gluing_nb}}
Let us explain how we computed the matrix $E_{S}$.
We first notice the following relations, that hold because $E_{m,m'}$ and $E_{m',m}$ are disjoint sets whenever $m\neq m'$:
$$
\left|E_{m,m'}\cup E_{m',m}\right|=
\begin{cases}
    \left(E_{S}\right)_{m,m'}\text{ if }m\neq m'
    \\
    \frac{1}{2}\left(E_{S}\right)_{m,m}\text{ if }m=m'
\end{cases}
$$

Second, the coefficients $\left|E_{m,m'}\cup E_{m',m}\right|$ can be interpreted as follows.
To build an ECTN $M\in E_{m,m'}$, we need to choose the profile $c$ of the central edge of $M$.
Let us denote by $i$ the central node of $m$ and by $j$ the central node of $m'$.
As the profile of $j$ in $m$ is the same as the profile of $i$ in $m'$, we should choose $c$ among the profiles common to $m$ and $m'$:
\begin{align*}
    \left|E_{m,m'}\cup E_{m',m}\right|
    &=
    \sum_{c\in\{a\in A|s(m,a)s(m',a)>0\}}
    \\
    &~
    B\left[\left(s(m,a)-\delta_{a,c}\right)_{a},
    \right.
    \\
    &~\left.
    \left(s(m',a)-\delta_{a,c}\right)_{a}\right]
\end{align*}

where $A$ denotes the set of possible NCTN profiles and $s(m,a)$ denotes the number of satellites with profile $a$ belonging to the NCTN $m$.

Before we explain what $B[.,.]$ stands for, let us precise that we will use in this sub-subsection and the next one, the map representation \ref{subsubsec:map_rep} for NCTN.
In this representation, a motif $m$ is actually a tuple of integers:
$$m\xleftrightarrow{}(s(m,a))_{a\in A}$$

Given two such tuples $n=(n_{a})_{a\in A}$ and $n'=(n'_{a})_{a\in A}$, $B[n,n']$ denotes the number of gluing matrices specifying which satellites of the NCTN whose map representation is $n$ match with which satellites of the NCTN whose map representation is $n'$.
More precisely, a gluing matrix between two tuples $n$ and $n'$ seen as representing NCTN, consists in a matrix $(b_{a,a'})_{a,a'\in A}$ where $b_{a,a'}$ denotes the number of satellites of $n$ with profile $a$ that match satellites of $n'$ with profile $a'$.
Formally, a gluing matrix between $n$ and $n'$ is defined by:
$$
\begin{cases}
    b_{a,a'}\in\mathbb{N}\\
    \sum_{a'}b_{a,a'}\leq n_{a}\\
    \sum_{a}b_{a,a'}\leq n'_{a'}
\end{cases}
$$

While the number of gluing matrices is difficult to compute in the general case, we can compute it for small numbers of possible profiles.
For example, if $|A|=1$, the above definition reduces to:

$$
\begin{cases}
    b_{1,1}\in\mathbb{N}\\
    b_{1,1}\leq n_{1}\\
    b_{1,1}\leq n'_{1}
\end{cases}
$$
This yields
$$
B_{1}(n,n')=1+\min(n,n'),\forall n,n'\in\mathbb{N}
$$
where we introduced a subscript on $B$ to explicit the number of components of $n$ and $n'$.
Here, we have a single component so we identified the tuple $n$ to its unique integer component.

For a generic $|A|$, the inequality $b_{a,a'}\leq\min(n_{a},n'_{a'})$ always holds.
This yields a polynomial upper bound on $B$:
$$
B_{|A|}(n,n')\leq\prod_{a,a'\in A}\left(1+\min(n_{a},n'_{a'})\right)
$$

However, taking this bound as approximation for $B$ results in an ECTN distribution that does not match at all the empirical case.
To determine whether this gap is due to the bad approximation or the hypothesis of maximum entropy, we need to compute $B_{|A|}(n,n')$ exactly.

To compute $B_{k}$ for a generic integer $k$, we will establish a recursion formula.
However, to speed up computations, it turns out that it is easier to compute the number of gluing matrices for $n$ and $n'$ having different sizes.

Indeed, it happens frequently that a motif $n$ has many components equal to zero.
If $n_{a}=0$, the condition involving $n_{a}$ for $b$ contributing to $B(n,m)$ yields
$b_{a,a'}=0,\forall a'$.
Said otherwise, the component $n_{a}$ can be removed from $n$ as long as the line of index $a$ is removed from $b$ too, turning $b$ into a non-square matrix.
Removing this way the zero components of $n$ and $m$ prior computing $B(n,m)$ speeds up the computations, but requires dealing with non-square matrices.

To do this, we define the set $\mathcal{B}_{k,l}(n,m)$ of gluing matrices between the integer tuples $n$ and $m$, where $n$ has $k$ components and $m$ has $l$ components:
\begin{align*}
\hspace{-1em}
    \mathcal{B}_{k,l}(n,m)
    &=
    \left\{
    b\in\mathcal{M}_{k,l}(\mathbb{N})\left|
    \sum_{a'}b_{a,a'}\leq n_{a},
    \sum_{a}b_{a,a'}\leq m_{a'}
    \right.\right\}
\end{align*}
where $\mathcal{M}_{k,l}(\mathbb{N})$ denotes the set of matrices with $k$ rows, $l$ columns and integer coefficients.

Then we can define the number of gluing matrices between $n$ and $m$:
\begin{align*}
B_{k,l}(n,m)
&=
\left|\mathcal{B}_{k,l}(n,m)\right|
\end{align*}

We can derive a recursion formula for $B_{k,l}(n,m)$, but before giving its expression, it is useful to introduce the following notation:
$$
\begin{cases}
    b_{.,+}=\left(\sum_{a'}b_{a,a'}\right)_{a}
    \\
    b_{+,.}=\left(\sum_{a}b_{a,a'}\right)_{a'}
\end{cases}
$$
Said otherwise, $b_{.,+}$ (resp. $b_{+,.}$) denotes the vector obtained by summing the matrix $b$ over its columns (resp. rows).

If $n$ is a vector of size $k$, it is useful to introduce $n[:i]$, which denotes the vector of size $k-i$ obtained by \textit{keeping} the first components of $n$.
Similarly, $n[i:]$ denotes the vector of size $k-i$ obtained by \textit{removing} the first components of $n$.

This makes the recursion formula for $B_{k,l}(n,m)$ more human readable:

\begin{align}
B_{k,l}(n,m)
&=
\sum_{b\in\mathcal{B}_{i,l-j}(n,m)}
\sum_{b'\in\mathcal{B}_{k-i,j}(n,m)}
\nonumber\\
&
B_{i,j}\left[n[:i]-b_{.,+},m[:j]-b'_{+,.}\right]
\nonumber\\
&
B_{k-i,l-j}\left[n[i:]-b'_{.,+},m[j:]-b_{+,.}\right]
\label{eq:rec_gluing_mat}
\end{align}

We prove in Appendix \ref{sec:AppendixA} that this formula holds for any integers $i,j,k,l$ as long as $i<k$ and $j<l$.

In our numerical implementation, we computed $B_{k,l}(n,m)$ recursively by choosing $i=\lfloor{\frac{k}{2}}\rfloor$ and $j=\lfloor{\frac{l}{2}}\rfloor$.
Moreover, we stored the values of $B_{k,l}(n,m)$ already computed.
This speeds up computations greatly, making possible the construction of the ECTN distribution in less than one minute on any modern laptop for motifs of depth 3 and temporal networks of the same size as the largest network currently available in the SocioPatterns collection.

\subsubsection{Recap of the steps leading to the MEP distribution}
Before we move on to the next section, let us make a pause to clarify how the ECTN distribution of maximum entropy -the MEP distribution- is computed from the NCTN distribution in practice.

The first step consists in computing the coefficients of the equation \ref{eq:lag_to_solve}, where the unknowns are the $\tilde{\gamma}_{m}$.
The right-hand side of this equation is computed from the NCTN distribution:
$V_{m}=N_{n}(m)|m|$.
Computing the coefficients $E_{S}$ appearing in the left-hand side is trickier:
\begin{align*}
    (E_{S})_{m,m'}
    &=
    (1+\delta_{m,m'})
    \sum_{c\in\{a\in A|m_{a}m'_{a}>0\}}
    \\
    &~
    B_{|A|}\left[\left(m_{a}-\delta_{a,c}\right)_{a},
    \left(m'_{a}-\delta_{a,c}\right)_{a}\right]
\end{align*}
where $m_{a}=s(m,a)$ and $m'_{a}=s(m',a)$.

For any couple $(n,m)$ of integer-valued vectors, the number of gluing matrices $B_{|A|}(n,m)$ is computed as follows:
\begin{enumerate}
    \item the zero components of $n$ and $m$ are removed.
    For example, if $n$ (resp. $m$) has $|A|-k$ (resp. $|A|-l$) components equal to zero, and if we denote by $n'$ (resp. $m'$) the vector with the zero components removed, we have:
    $B_{|A|}(n,m)=B_{k,l}(n',m')$
    \item if $k,l\leq2$, return the value of $B_{k,l}(n',m')$ from a recursion-free formula.
    Else, check whether $B_{k,l}(n',m')$ has already been computed.
    If yes, return its value stored in the cache.
    If no, compute it by using the recursion formula \ref{eq:rec_gluing_mat} with $i=\lfloor{\frac{k}{2}}\rfloor$ and $j=\lfloor{\frac{l}{2}}\rfloor$.
    \item store the computed value in the cache and return to step 1. to compute the terms produced by the recursion.
\end{enumerate}

At this point, we have computed both the coefficients $E_{S}$ and $V$ of eq \ref{eq:lag_to_solve}, so we can invert it to deduce the $\tilde{\gamma}$.
This is done by using the numerical solver scipy.optimize.least\_squares with the following input:
$$
f(\tilde{\gamma})=
\tilde{\gamma}\cdot(E_{S}\tilde{\gamma})-V
$$
The numerical solver will minimize the cost function:
$$
L(\tilde{\gamma})=
\sum_{m}f(\tilde{\gamma}_{m})^{2}=
||\tilde{\gamma}\cdot(E_{S}\tilde{\gamma})-V||_{2}^{2}
$$

Once the $\tilde{\gamma}$ have been computed, the MEP distribution is deduced from eq \ref{eq:ectn_lag}, that we recall here for completeness:
$$
N_{e}(M)=
\frac{1}{d}\tilde{\gamma}_{\psi_{1}(M)}\tilde{\gamma}_{\psi_{2}(M)}    
$$

\section{probing spatial correlations with motifs}
We have introduced temporal motifs that include triangles.
We call them edge-center temporal neighborhoods motifs (ECTN) and we showed how they were related to the node-centered motifs (NCTN) of the literature.

Now, in a modeling perspective, we would like to estimate how complex a generative model should be in order to reproduce a given ECTN distribution.
More precisely, we would like to identify what minimal mechanisms can account for the ECTN distribution sampled from a given data set.
To achieve this, the idea is to assume some hypothesis on the dynamics generating the temporal network.
Then we derive analytically the ECTN distribution we should observe according to this hypothesis.
In this section, we will consider three distinct hypotheses, of increasing complexity, under which we derive an analytical expression for the ECTN distribution.

\subsection{Hypothesis 1: independent edges}
The simplest process that generates a temporal network has no spatial correlations, meaning that each edge of the network has its own independent dynamic.
We refer to this hypothesis as the hypothesis of independent edges.
Before we derive the ECTN distribution under this hypothesis, let us tackle the NCTN distribution.

\subsubsection{The NCTN distribution in the case of independent edges}
Using the map representation, a NCTN motif can be written as
$$\vec{n}=\left(n_{a}\right)_{a\in A}$$
where $n_{a}$ = number of satellites with activity profile $a$, and $A$ = set of possible profiles.

As we exclude the empty motif in the mining phase, we are actually interested in the probability
$$P(\vec{n}|\vec{n}\neq\vec{0})=\frac{P(\vec{n})}{1-P(\vec{0})}$$
that we will denote as $P^{*}(\vec{n})$.

In the setting of independent edges, choosing a NCTN motif amounts to choosing the number of satellites, then assigning those satellites activity profiles.
This writes as:
$$P^{*}(\vec{n})=\Omega(n)n!\prod_{a\in A}\frac{p_{a}^{n_{a}}}{n_{a}!}$$
where $n=\sum_{a\in A}n_{a}$ = number of satellites, $\Omega(n)$ = probability of having $n\geq1$ satellites and $p_{a}$ = probability for an edge to have $a$ as activity profile.
Thus $\sum_{a\in A}p_{a}=1$.

Note that $\Omega$ is nothing but the distribution of the node degree, if we exclude the degree zero and aggregate the network on a sliding window of size equal to the depth of the considered motifs.

\subsubsection{The ECTN distribution in the case of independent edges}
Let us compute the ECTN distribution in the same setting as above.

Consider two nodes $i$ and $j$, a time $t$ and a depth $d$.
We are interested in the probability that the ECTN instance $((i,j),t,d)$ is an instance of $M$, where $M$ is a fixed ECTN motif.

Like in the NCTN case, we restrict our distributions to non-empty motifs.
Let us refer to our motif $M$ as $(c,\vec{n})$, where $c$ is the profile of the central edge and $n_{a}$ is the number of satellites with profile $a$.
Because of the distinction between the two central nodes, components of $n$ are not independent from each other.

To leverage this issue, we will compute first the probability $P$ of $(c,\vec{n})$ for any $\vec{n}$, defined by arbitrarily associating the letter `1' to the node $i$ and the letter `2' to the node $j$.
Then we deduce the probability $P^{*}$ of observing the motif $M$:
\begin{equation} \label{eq:0}
    P^{*}(M)=
    \begin{cases}
        P(M)\text{ if the central nodes are equivalent}\\
        2P(M)\text{ else}
    \end{cases}
\end{equation}

The case of equivalence between the central nodes corresponds to the case where the two strings from Figure \ref{fig:string_rep_ECTN} produced by either naming choice are equal.

This being said, we can write down the probability of the couple $(c,\vec{n})$:
$$P(c,\vec{n})=P^{*}(c)\Omega(n)n!\prod_{a\in A}\frac{p_{a}^{n_{a}}}{n_{a}!}$$
Note that the distribution $\Omega$ and the set $A$ are not the same as in the NCTN case.
In particular, $\Omega(n)$ denotes the probability that a pair of nodes interacts with $n$ other nodes in a time window of size $d$, the motif depth.
The value $n=0$ is now allowed, as we collect any ECTN that have a non-empty central profile $c\neq\emptyset$.
However, we still have $n=\sum_{a}n_{a}$ and $\sum_{a}p_{a}=1$.

As edges are mutually independent, our formula can be expanded further.
Indeed, a profile of a satellite in an ECTN is actually a pair of edge profiles.
For example, the profile `123' of the satellite $k$ comprise the edge profile `101' for the edge $(i,k)$ and `011' for the edge $(j,k)$.
Also, the profile `111' comprise the edge profiles `111' and `000'.
More generally, if $a$ is the profile of an ECTN satellite, the two edge profiles it involves are $\phi_{1}(a)$ and $\phi_{2}(a)$.
The $\phi_{i}$ are maps defined as follows:
$\phi_{1}$ replaces every `2' in a string by a `0' and any `3' by a `1', every other letter being unchanged.
$\phi_{2}$ replaces every `1' in a string by a `0', any `2' by a `1' and any `3' by a `1', any other letter being unchanged.

Then, considering $Q_{a}$ the probability for an edge to have $a$ as activity profile, \textit{empty profile included}, we can safely write that
$$p_{a}=P(a|a\neq\emptyset)=\frac{Q_{\phi_{1}(a)}Q_{\phi_{2}(a)}}{1-P(a=\emptyset)}=\frac{Q_{\phi_{1}(a)}Q_{\phi_{2}(a)}}{1-Q_{\emptyset}^{2}}$$

We deduce the expression for $P(M)$:
\begin{equation}\label{eq:13}
    \hspace*{-1cm}P(c,\vec{n})=
    \frac{Q_{c}}{1-Q_{\emptyset}}\Omega(n)\left(1-Q_{\emptyset}^{2}\right)^{-n}n!\prod_{a\in A}\frac{\left(Q_{\phi_{1}(a)}Q_{\phi_{2}(a)}\right)^{n_{a}}}{n_{a}!}
\end{equation}

Together with equation \ref{eq:0}, it allows to compute $P^{*}(M)$ for any ECTN $M$.

\subsection{Hypothesis 2: conditional independence of edges}

Assuming independent edges makes triangles unlikely to occur.
To have a glimpse of this, let us consider an ECTN profile $a$ such that $\phi_{1}(a)\neq\emptyset$ and $\phi_{2}(a)=\emptyset$.
Then we have $p_{a}\propto Q_{\phi_{1}(a)}Q_{\emptyset}$.
If now we replace every `1' in $a$ by a `3', we get the profile $a'$ such that
$p_{a'}\propto Q_{\phi_{1}(a)}^{2}$

We deduce
$$
\frac{p_{a'}}{p_{a}}=\frac{Q_{\phi_{1}(a)}}{Q_{\emptyset}}\ll1
$$
The last inequality holds in the empirical data sets we consider, because of the sparsity of social interactions.
Despite this sparsity, ECTN with triangles are much more likely than predicted when assuming independent edges.
This motivates for another hypothesis, under which triangles are more likely to occur.
For a triangle to occur, there must exist a correlation between adjacent edges.
The simplest way to introduce such a correlation while keeping the computation of the ECTN distribution tractable, is to assume that nodes are independent from each other.

To give a meaning to this idea, we associate a hidden state $h_{i}$ to each node $i$ of the network.
The independence condition means that the variables $h_{i}(t)$ and $h_{j}(t')$ are statistically independent whenever $i\neq j$.
Besides these hidden states, we assume the existence of a measurement process, that takes the hidden state as input and returns the interaction graph between nodes at the current time.
This modeling is akin to a hidden Markov model that would generate the temporal network we observe.
Examples of such models in the literature are \cite{boguna2003class} for growing a static network and \cite{hartle2021dynamic} for generating a temporal network.
As we aim at starting simple, let us assume that at each time $t$, the interaction graph $G(t)$ is generated as follows:
$$
\forall i<j, P\left((ij)\in G_{1}(t)|h(t)\right)=f(h_{i}(t),h_{j}(t))$$
with $G_{1}(t)$ denoting the set of edges of $G(t)$ and $f$ an arbitrary function with values in $[0,1]$.

Note that in this model, edges are independent when conditioned to the current state $h$, but are not otherwise.
However, if we consider that the hidden states are certain random variables (i.e. they can only take a single value), then edges become independent from each other, but possibly have different probabilities to occur.

In the following, we will assume that the variables $h_{i}(t)$ are independent, real-valued, identically distributed according to a density probability $\rho(h)$ and not certain.

Under these hypotheses, we will compute the ECTN distribution for motifs of generic depth.

Let us consider two nodes $i$ and $j$ that will constitute the central nodes of our ECTN.
Let us denote the motif depth by $d$.
Then, using the map representation, we can write the ECTN as a tuple $(c,(n_{a})_{a\in A_{e}(d)})$ where $c\in A_{n}(d)$ denotes the profile of the central edge $(i,j)$, $A_{n}(d)$ (resp. $A_{e}(d)$) denotes the set of possible NCTN (resp. ECTN) profiles for satellites.

However, the map representation is not suitable to the computation of the ECTN distribution.
Indeed, our hypothesis states that different hidden variables are attributed to different nodes.
So to compute the probability that the ECTN $M$ is associated to the edge $(i,j)$, we need to know which central node ($i$ or $j$) corresponds to the letter `1' or `2'.
Otherwise, we will be unable to link the ECTN $(c,\vec{n})$ to interactions with either $i$ or $j$.

To solve this, we will represent an ECTN by an unordered pair of equivalent tuples:
the tuple $(c,\vec{n}^{(i)})$ where the letter `1' is associated to the node $i$ and the tuple $(c,\vec{n}^{(j)})$, where `1' is associated to $j$.

Note that the central profile is independent from the choice of dominant node, and that some ECTN are invariant under the change of convention, i.e. $\vec{n}^{(i)}=\vec{n}^{(j)}$.
These are the ECTN whose graphical representation is invariant under a vertical flip.
To avoid redundancy for symmetric ECTN, we will represent the ECTN centered on the edge $(i,j)$ by the following tuple:
$$
\begin{cases}
    (c,[\vec{n}])
    \\
    [\vec{n}]=\left\{\vec{n}^{(i)},\vec{n}^{(j)}\right\}
\end{cases}
$$

Then we can write the probability for such an ECTN to occur as $P(c,[\vec{n}])$.
We have:

\begin{align*}
    P(c,[\vec{n}])
    &=
    \sum_{h_{1},\hdots,h_{N}}P(c,[\vec{n}]|\vec{h})\prod_{i=1}^{N}\rho(h_{i})
\end{align*}
where
$$
\begin{cases}
    \vec{h}=(h_{1},\hdots,h_{N})
    \\
    P(c,[\vec{n}]|\vec{h})=
    \sum_{\vec{n}\in[\vec{n}]}
    P(c,\vec{n}|\vec{h})
\end{cases}
$$

Now, it turns out that we can compute $P(c,\vec{n}|\vec{h})$.
In the Appendix \ref{sec:AppendixD}), we show that
\begin{align*}
    P(c,\vec{n}|h_{i},h_{j})
    &\simeq
    f_{c}(c|h_{i},h_{j})\Omega(n|h_{i},h_{j})n!
    \prod_{a\in A}\frac{p_{a}(h_{i},h_{j})^{n_{a}}}{n_{a}!}
\end{align*}
where $n=\sum_{a\in A_{e}(d)}n_{a}$ and we have introduced
$$
p_{a}(h_{i},h_{j})=\frac{\left<\tilde{f}_{a}(h)\right>}{\left<f_{\Omega}(h)\right>}
$$

The complete expression for the ECTN probability writes:
\begin{align*}
    P(c,[\vec{n}])
    &=
    \binom{n}{\vec{n}}\sum_{h,h'}\rho(h)\rho(h')f_{c}(c|h,h')
    \Omega(n|h,h')
    \\
    &~
    \sum_{\vec{n}\in[\vec{n}]}
    \prod_{a\in A_{e}(d)}
    p_{a}(h,h')^{n_{a}}
\end{align*}
Excluding the ECTN with an empty profile for the central edge leads to:
$$
P^{*}(c,[\vec{n}])=
\frac{P(c,[\vec{n}])}{1-\sum_{h,h'}\rho(h)\rho(h')f_{c}(\emptyset|h,h')}
$$

\subsubsection{Testing the hypothesis of independent nodes}
\label{subsubsec:cond_ind_test}
We know how to compute the ECTN distribution assuming either independent edges or independent nodes.
Now, we would like to compare this theoretical distribution with an empirical distribution sampled from a given temporal network.
In the case of the hypothesis of independent edges, this comparison is easy:
we only have to sample $Q_{a}$ and $\Omega(n)$.

However, testing the hypothesis of independent nodes requires inferring the distribution of the hidden states $\rho$ and the measurement process $f$.
This would result in a model with an infinite number of parameters, as soon as the number of different realizations for $h$ is infinite.
Then this model may easily overfit an empirical ECTN distribution.
Hence, before inferring any hidden parameter, we will check that not any distribution can be written under the form that arises from assuming independent nodes.

Under this assumption, the ECTN distribution has the following formal structure:
\begin{equation*}
\begin{cases}
    P(c,\vec{n})
    =
    \binom{n}{\vec{n}}
    \sum_{e}
    f_{e}(c)\Omega_{e}(n)
    \prod_{k=1}^{D}p_{e,k}^{n_{k}}
    \\
    f_{e}(c),\Omega_{e}(n),p_{e,k}
    \geq0,\forall e,c,n,k
    \\
    \sum_{k}p_{e,k}
    =
    1,\forall e
\end{cases}
\end{equation*}
where $n=\sum_{k=1}^{D}n_{k}$ and $D$ is an integer equal to the number of components of $\vec{n}$ (using the previous notations, it would equal to $|A_{e}(d)|$).

Writing $P(c,\vec{n})$ this way makes clear that assuming independent nodes is a particular case of assuming independent edges when conditioned on some hidden variables (indexed here by $e$).
Indeed, each term has the same form as eq \ref{eq:13}, which gives the expression of $P(c,\vec{n})$ in the case of independent edges.

In practice, we would like to test directly the hypothesis of \textit{conditional independence of edges}, where the unknown parameters are the $f_{e}(c),\Omega_{e}(n),p_{e,k}$; instead of testing the hypothesis of \textit{independent nodes}, where the unknowns are the $f(h,h'),\rho(h)$.

\subsubsection{Testing the hypothesis of conditional independence}
To check that the hypothesis of conditional independence allows to discriminate between distinct classes of temporal networks, we prove in Appendix \ref{sec:cond_ind_test} that not every distribution on tuples of integers $g(\vec{n})$ can be approximated by the following form:
\begin{equation}\label{eq:10}
\begin{cases}
    g(\vec{n})
    =
    \binom{n}{\vec{n}}
    \sum_{e}
    \Omega_{e}(n)\prod_{k=1}^{D}p_{e,k}^{n_{k}}
    \\
    \Omega_{e}(n),p_{e,k}
    \geq0,\forall e,n,k
    \\
    \sum_{k}p_{e,k}
    =
    1,\forall e
\end{cases}
\end{equation}
where $D\geq2$, $\vec{n}=(n_{1},\hdots,n_{D})$ and $n=\sum_{k}n_{k}$.

Since we have -at least in the case of a countable number of hidden variables $e$- the certainty that not every distribution can be approximated by this generic form, we deduce that testing for the hypothesis of conditional independence of edges rather than the independence of nodes makes sense:
we can look for $f_{e}$, $\Omega_{e}$ and $p_{e,k}$ that minimize the following cost function:
\begin{equation}\label{eq:cond_ind}
    L(f,\Omega,p)=
    \sum_{M}\left(P^{\text{th}}(M;f,\Omega,p)-P^{\text{xp}}(M)\right)^{2}
\end{equation}
where $P^{\text{xp}}$ denotes the ECTN distribution we want to fit:
given a temporal network, $P^{\text{xp}}(M)$ is the number of instances of the ECTN $M$ in this network, then normalized such that $\sum_{M}P^{\text{xp}}(M)=1$.

For completeness, let us explicit the expression of $P^{\text{th}}$:
\begin{align*}
    P^{\text{th}}(c,[\vec{n}];f,\Omega,p)
    &=
    \binom{n}{\vec{n}}
    \sum_{e}f_{e}(c)\Omega_{e}(n)
    \sum_{\vec{n}\in[\vec{n}]}
    \prod_{k=1}^{D}p_{e,k}^{n_{k}}
\end{align*}

The minimization of $L$ is subject to the following constraints:
$$
\begin{cases}
    f_{e}(c),\Omega_{e}(n),p_{e,k}
    \geq0,\forall e,c,n,k
    \\
    \sum_{k}p_{e,k}
    =
    1,\forall e
    \\
    D=|\vec{n}|
\end{cases}
$$
Note that contrary to $n$, the variable $c$ can only take a finite number of values, equal to $|A_{n}(d)|$.
In practice, we restrict $n$ to the values encountered in the ECTN distribution we want to fit.

\section{edge-centered motifs in empirical networks and models: comparison with the theory}
Let us apply the theory we have developed to some empirical temporal networks.
First, we will describe the networks under study, second we will investigate the principle of maximum entropy and third we will investigate the hypotheses of independent edges and conditional independence of edges.

\subsection{\label{subsec:2}Temporal network data sets}

We consider 27 data sets describing temporal networks, corresponding to
(i) 14 publicly available empirical data sets on social interactions with high temporal resolution
\cite{barrat2013temporal,sociopatterns,toth2015role} and
(ii) 13 models of temporal networks.
We consider 6 models representing the dynamics of pedestrians and their physical proximity, 4 temporal network models from the Activity Driven with Memory (ADM) class \cite{perra2012activity,Laurent_2015,le2023modeling}, and 3 ad hoc models of temporal edge dynamics.

In this sub-section, we give a brief description of the data sets:
for the empirical data, we indicate when and where the data were collected and for the models we give the principle behind them.
For more details about the sizes of the different data sets, see Appendix \ref{sec:AppendixC}.
For more details about the models, see the supplementary material of \cite{le2024flow}.

Note that, while here for illustration purposes we focus on temporal networks of face-to-face interaction, our framework is applicable to any type of temporal networks, including networks of higher-order interactions \cite{battiston2020networks}.

\subsubsection{Empirical data sets}

The empirical temporal networks we use represent face-to-face interaction data collected in various contexts using wearable sensors that exchange low-power radio signals
\cite{sociopatterns,toth2015role}. 
This allows to detect face-to-face close 
proximity with here a temporal resolution of about 20 seconds \cite{cattuto2010dynamics}. All the data we used have been made publicly available by the research collaborations who collected the data \cite{sociopatterns,toth2015role}. They correspond to 
data collected among human individuals in conferences, schools, a hospital, workplaces, and also within a group of baboons.
In all cases, 
individuals are represented as nodes, and an edge is drawn between two nodes each time the associated individuals are interacting with each other. 

The data sets we consider are:
\begin{itemize}
    \item ``conf16'', ``conf17'', ``conf18'', ``conf19'': these data sets were collected in scientific conferences, respectively 
    the 3rd GESIS Computational Social Science Winter Symposium (November 30 and December 1, 2016), 
    the International Conference on Computational Social Science (July 10 to 13, 2017),
    the Eurosymposium on Computational Social Science (December 5 to 7, 2018), and the 41st European Conference on Information Retrieval (April 14 to 18, 2019) \cite{genois2022combining};
    
   \item the ``utah'' data set describes the proximity interactions which occurred on November 28 and 29, 2012 in an urban public middle school in Utah (USA) \cite{toth2015role};
    \item the ``french'' data set contains the temporal network of contacts between the children and teachers that occurred in a french primary school on Thursday, October 1st and Friday, October 2nd 2009. It is described in \cite{Gemmetto2014,10.1371/journal.pone.0023176};   
    \item the ``highschool1'', ``highschool2'' and 
    ``highschool3'' data set describe the interactions between students in a high school in Marseille, France 
\cite{10.1371/journal.pone.0107878,mastrandrea2015contact}. They were respectively collected for
 three classes during four days in Dec. 2011, 
 five classes during seven days in Nov. 2012 
 and nine classes during 5 days in December 2013;

    \item the ``hospital'' data set contains the temporal network of contacts between patients and health-care workers (HCWs) and among HCWs in a hospital ward in Lyon, France, from December 6 to 10, 2010. The study included 46 HCWs and 29 patients \cite{10.1371/journal.pone.0073970};

    \item the ``malawi'' data set contains the list of contacts measured between members of 5 households of rural Kenya between April 24  and May 12, 2012 \cite{kiti2016quantifying};

    \item the ``baboon'' data set contains observational and wearable sensors data collected in a group of 20 Guinea baboons living in an enclosure of a Primate Center in France, between June 13 and July 10 2019 \cite{gelardi2020measuring};
    
    \item the ``work1'' and `work2''
    data sets contain the temporal network of contacts between individuals measured in an office building in France, respectively from June 24 to July 3, 2013 \cite{NWS:9950811} and 
     during two weeks in 2015 \cite{Genois2018}.
\end{itemize}

\subsubsection{Pedestrian models}

Pedestrian models consist in stochastic agent-based models implemented in discrete time.
These agents move through a two-dimensional space and are point oriented particles.
In the simulations considered here, the 2D space is a square with reflecting boundary conditions.
A temporal network is built from the agents' trajectories according to a rule similar to the one used in empirical face-to-face interactions:
an interaction between two agents $i$ and $j$ is recorded at time $t$ if $i$ and $j$ are close enough and oriented towards each other at $t$.

In all pedestrian models considered in this paper, agents are point particles.
Three types of models are considered:
\begin{itemize}
    \item Brownian particles without any interaction:
    ``brownD01'' and ``brownD001''.
    These models differ only by the value of the diffusion coefficient of the agents.
    \item active Brownian particles \cite{solon2015active}:
    ``ABP2pi'' and ``ABPpi4''.
    The orientation vector of an active Brownian particle follows a Brownian motion, whereas its position vector enjoys an overdamped Langevin equation with a self-propelling force.
    This force has constant magnitude and is parallel to the orientation vector.
    In our case, however, the noise contributing to the velocity vector is zero, meaning the velocity is equal to the self-propelling force.
    Besides, the noise giving the angular velocity is a uniform random variable in $[-\theta,\theta]$.
    In ``ABP2pi'', $\theta=\pi$ and in ``ABPpi4'', $\theta=\frac{\pi}{8}$.
    \item the Vicsek model \cite{vicsek1995novel}:
    ``Vicsek2pi'' and ``Vicsekpi4''.
    In this model, the velocity of a particle at the next time step $t+1$ points in the same direction as the average velocity of its neighbours at time $t$, and an angular noise is added.
    The velocity modulus is constant and identical for every particle.
    In ``Vicsek2pi'', the velocity direction is drawn uniformly in an interval of size $2\pi$ around the average velocity of the neighbours, i.e., is completely random. In ``Vicsekpi4'', the velocity direction is drawn uniformly in an interval of size $\frac{\pi}{4}$ around the average velocity of the neighbours.
\end{itemize}

We report in Appendix \ref{sec:AppendixC} the sizes and durations used in each model.

\subsubsection{\label{subsubsec:4}Activity Driven with Memory models}
The class of Activity Driven with Memory (ADM) models is an extended framework \cite{le2023modeling} of the original Activity Driven (AD) model \cite{perra2012activity}.
In practice, an ADM model is a stochastic agent-based model in discrete time that produces a synthetic temporal network of interactions between agents.

We refer to \cite{le2023modeling} and the supplementary material of \cite{le2024flow} for a detailed description of the models and their phenomenology.
Here, we provide a brief reminder of their definition. 
In a nutshell, we consider $N$ agents, each endowed with an intrinsic activity parameter, and who interact with each other at each discrete time step in a way depending on their activity and on the memory of past interactions between agents.
This memory is encoded in another temporal network between the same agents, called the social bond graph:
in this weighted and directed temporal graph, the weight of an edge represents the social affinity of an agent towards an other agent.
At each time step, agents thus choose partners to interact with depending on their social affinity towards other agents. The affinity is then updated by the chosen interactions through a reinforcement process:
social bonds between interacting agents strengthen while the social affinity weakens if two agents do not interact.

The ADM models we considered in this paper are ``ADM9conf16'', ``ADM18conf16'', ``min\_ADM1'' and ``min\_ADM2''. Here
the numbers ``9'' or ``18'' stand for the specific dynamical rule of the model (in \cite{le2023modeling} a large number of possible variations of ADM rules has been explored) and the suffix ``conf16'' means that the parameters of the model have been tuned in order to resemble the empirical data set ``conf16''.
We report in Appendix \ref{sec:AppendixC} the sizes and durations used for each model.

\subsubsection{Edge-weight models}

An edge-weight (EW) model is also a stochastic agent-based model in discrete time producing a temporal network. However, contrary to the ADM or pedestrian models, here agents are not nodes but edges of this temporal network.

In these models, edges are independent of each other.
Their probability of activation is given by the fraction of time they have been active since their last ``birth'', which is defined either as the starting time of the temporal network, i.e. the time step 0, or as the last time the edge's history has been reset. We refer to the supplementary material of \cite{le2024flow} for more details on the three variants we consider here, denoted ``min\_EW1'', ``min\_EW2'' and ``min\_EW3''.
See Appendix \ref{sec:AppendixC} for the sizes used in the simulations of the models.

\subsection{Testing the principle of maximum entropy (MEP)}
\subsubsection{Approach by direct computation of the MEP distribution}

To test the principle of maximum entropy for a given ECTN distribution, we need first to extract the NCTN distribution.
Then we need to solve the equation \ref{eq:lag_to_solve} with a numerical solver (scipy.optimize.least\_squares) and inject the solution in eq \ref{eq:ectn_lag} to determine the MEP probability of each observed ECTN.

However, solving eq \ref{eq:lag_to_solve} can be time consuming when the number of distinct observed NCTN is large ($\sim 400$).
To reduce the computational load, we remove the less frequent motifs from the NCTN distribution.
To do this, we introduce a parameter $r$ and remove every motif $m$ that satisfies
$$
\frac{N_{n}(m)}{N_{n}^{\text{min}}}<r
$$
where $N_{n}^{\text{min}}=\min_{m}N_{n}(m)$

In practice, we also truncate the ECTN distribution to evaluate using the same parameter.
Thus, we compute the MEP probability of the ECTN $M$ that satisfy
$$
\frac{N_{e}(M)}{N_{e}^{\text{min}}}>r
$$
where $N_{e}(M)$ is the observed number of occurrences of $M$.

In the following, we take $r=20$ but we checked the consistency of our results with smaller $r$ (not shown).

Let us explain how we evaluate the quality of a given hypothesis (here the hypothesis of maximum entropy) for a given temporal network.
To measure how compatible is an hypothesis with a given motif distribution, that we will denote as $P^{\text{xp}}$ (see eq \ref{eq:cond_ind} for definition), we compute the following quantity:
$$
C=
1-\max_{M,P^{\text{xp}}(M)>0}
\frac{\left|P^{\text{xp}}(M)-P^{\text{th}}(M)\right|}{P^{\text{xp}}(M)+P^{\text{th}}(M)}
$$
where $P^{\text{th}}$ is the motif distribution derived from our hypothesis.

The compatibility coefficient $C$ is bounded between 0 and 1.
The maximum compatibility is achieved when the predicted and observed probabilities coincide for every observed ECTN, in which case $C=1$.

On the contrary, if for one observed motif $M$ we have that $P^{\text{th}}(M)=0$, then $C=0$ even if every other motif is perfectly explained by the tested hypothesis.
Thus, to perform well, an hypothesis needs to be compatible with \textit{every} observed motif.

To visualize how compatible each temporal network we consider is with a given hypothesis, we represent a temporal network by a point with coordinates $(C_{2},C_{3})$, where $C_{d}$ denotes the compatibility of motifs of depth $d$.

To discriminate between low-performing data sets, we also use a logarithmic scale, i.e. the coordinates are $(\log_{10}(C_{2}),\log_{10}(C_{3}))$.

This procedure for quantifying and visualizing how compatible is an hypothesis with our data sets will be used in the next subsections, to study the hypotheses of independent edges and conditional independence.

\begin{figure*}
    \subfigure[linear scale]{
        \includegraphics[width=\columnwidth]{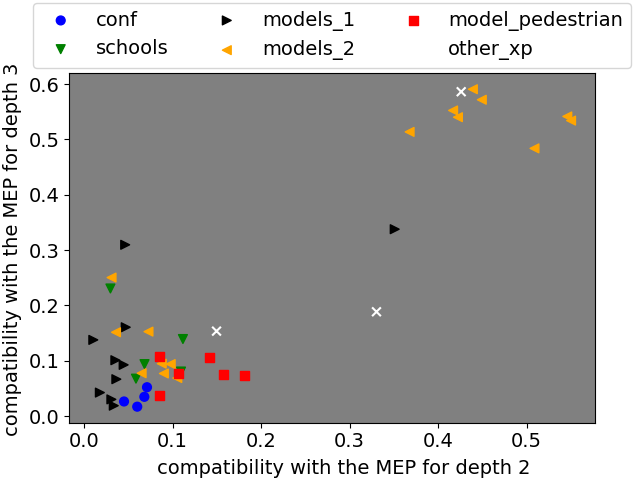}
    }
    \subfigure[logarithmic scale]{
        \includegraphics[width=\columnwidth]{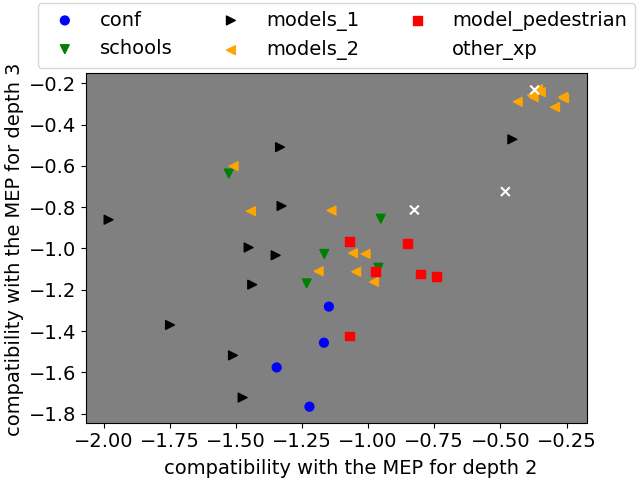}
    }
    \caption{\textbf{Compatibility of various data sets with the principle of maximum entropy.}
    Each data set is associated a point of coordinates $(C_{2},C_{3})$, where $C_{d}$ is the compatibility of the distribution of ECTN of depth $d$ with the MEP hypothesis.
    (a) linear scale:
    most models are clearly not compatible with the MEP.
    Interestingly, the compatible models stem from the family ``models\_2'', where the temporal network is generated by combining NCTN/ETN motifs.
    (b) logarithmic scale:
    6 empirical data sets have a compatibility less than 0.1 for both depths, but a few have a compatibility higher than this threshold.
    Overall, it is difficult from this figure alone to conclude whether the MEP hypothesis is a useful description of empirical networks.
    }\label{fig:MEP_hyp_compa}
\end{figure*}

We display on Figure \ref{fig:MEP_hyp_compa} the various data sets on the compatibility plane $(C_{2},C_{3})$ for the hypothesis of maximum entropy.

From this figure, we see that the least compatible data sets include many models whereas a few empirical networks have a compatibility higher than 0.1 for both depths.
However, this compatibility remains lower than $0.1$ for both depths for 6 empirical data sets.
Thus, the MEP hypothesis is not enough to fully explain the ECTN distribution.
Hence, we will try to identify which motifs are responsible for the compatibility gap.

To do this, given a distribution $P^{\text{xp}}$ and a motif $M$, we define its log-compatibility:
$$
\text{LC}(M,P^{\text{xp}})=\log_{10}\left(1-\frac{\left|P^{\text{xp}}(M)-P^{\text{th}}(M)\right|}{P^{\text{xp}}(M)+P^{\text{th}}(M)}\right)
$$

\begin{figure*}
    \subfigure[depth 2]{
        \includegraphics[width=\columnwidth]{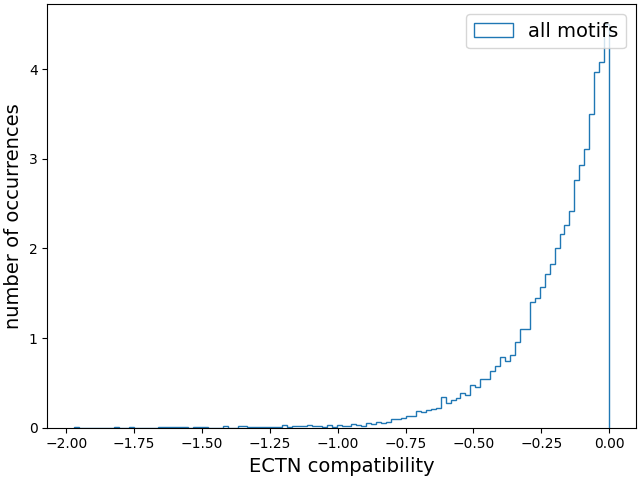}
    }
    \subfigure[depth 3]{
        \includegraphics[width=\columnwidth]{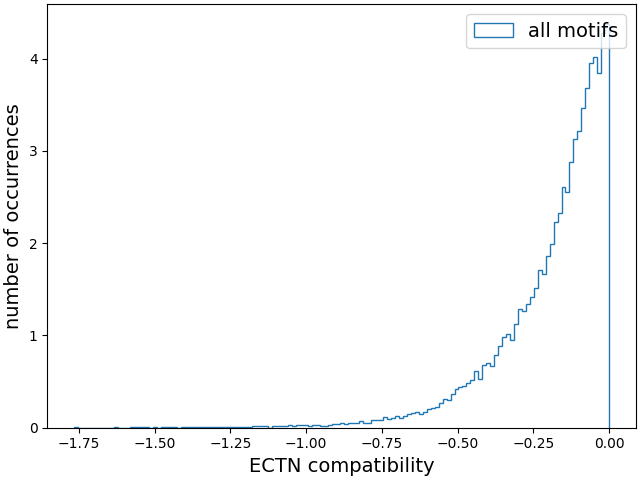}
    }
    \caption{\textbf{Identifying the motifs of lower compatibility.}
    What is shown are the histograms of the ECTN log-compatibility, sampled on every ECTN and data set.
    A single mode is visible, meaning that the compatibility of a motif cannot be predicted from its string sequence (cf text).
    (a) ECTN of depth 2.
    (b) ECTN of depth 3.
    }\label{fig:low_compa_MEP}
\end{figure*}

Sampling LC for every motif and data set yields the distributions visible on Figure \ref{fig:low_compa_MEP}, that we call compatibility histograms.

A single mode is visible in these histograms, so drawing them fails to identify why some motifs do not fit to the principle of maximum entropy.
Actually, it tells us that the compatibility of a motif cannot be predicted from its string sequence.
Indeed, suppose we have a mapping
$M\mapsto \text{LC}(M)$ satisfying
$$
\text{LC}(M)=\text{LC}(M')\text{ iif }f(M)=f(M')
$$
where $f$ is independent of $P^{\text{xp}}$.
For example, $f(M)$ is equal to the number of triangles in $M$.

Then the compatibility values would separate according to the values of $f$, so we would observe as many modes in the compatibility histogram as $f$ has distinct images.

As a single mode is visible on Figure \ref{fig:low_compa_MEP}, we deduce that the mismatch between some motifs and the principle of maximum entropy is not due to the expression of these motifs.
This means that \textit{a priori} any motif could fit or deviate from the principle of maximum entropy.
Then, to understand why the MEP hypothesis fails to grasp some motifs, but globally seems to perform well, we need another approach.
Said otherwise, we will try to solve the same equation \ref{eq:lag_to_solve} as before, but using an alternative to the numerical solver scipy.optimize.least\_squares.

\subsubsection{Approach by completion of the MEP distribution (partial computation)}
The difficulty in understanding how well the principle of maximum entropy describes empirical data, is that the parameters $\tilde{\gamma}_{m}$ are output of a black box process, namely the numerical solving of equation \ref{eq:lag_to_solve}.

Recall that the ECTN distribution will satisfy the principle of maximum entropy as soon as it writes under the form
\begin{equation}\label{eq:4}
    N_{e}(M)=\gamma(\psi_{1}(M))\gamma(\psi_{2}(M))
\end{equation}
where $\gamma$ is a map taking as input a NCTN and returning a positive real number.

In this sub-subsection, instead of trying to compute the MEP distribution for every ECTN, we will assume that we already know a part of this distribution, and then show how to compute the remaining part.
More precisely, we separate between a set of ECTN that we use to compute the map $\gamma$ from eq \ref{eq:4} and the remaining ECTN, for which we test whether the equation \ref{eq:4} holds.
These two sets can be seen as a training and a test set:
the training set serves to determine the map $\gamma$ while the test set serves to check whether eq \ref{eq:4} is satisfied by the whole ECTN distribution.

This approach is less ambitious than the previous one, since it cannot be used to infer the whole ECTN distribution.
However, if some ECTN have already been sampled in a given network, it speeds up the computation of the whole distribution by using eq \ref{eq:4} to deduce the number of instances of the remaining ECTN.

The goal is to sample as less ECTN as possible (choose a minimal training set), and then deduce $N_{e}$ for the remaining ECTN.
One example of minimal training set is the following.
First, notice that for some ECTN $M$, we have $\psi_{1}(M)=\psi_{2}(M)$.
Those are the ECTN whose graphical representation is symmetric under a vertical flip, or whose string representation is invariant under the exchange of the letters `1' and `2' in the satellite profiles.

The shortest ECTN that have this property are the ECTN without any satellite.
For a depth $d$, there are $2^{d}-1$ such ECTN.
These ECTN are equivalent to their central profile:
$M=c(M)$.
We have
$$
\forall M,\gamma(c(M))=\sqrt{N_{e}(c(M))}
$$

Then, let us consider the ECTN whose string representation does not contain any letter `2' nor the letter `3'.
These one-sided ECTN are equivalent to their sub-NCTN $\psi_{1}(M)$ as well as the specification of their central profile $c(M)$.
So we have:
$$
\forall M,\gamma(\psi_{1}(M))=\frac{N_{e}(c(M),\psi_{1}(M))}{\sqrt{N_{e}(c(M))}}
$$

As $\psi_{1}$ is a surjective mapping from the space of ECTN to the space of NCTN, the map $\gamma$ has been fully specified.

To recap, we have used the family of one-sided ECTN as a training set allowing to fully specify the map $\gamma$.
Instead, we could have considered any set of ECTN whose image under $\psi_{1}$ yields all the possible NCTN.

Now, we would like to evaluate the compatibility of a given empirical distribution $N_{e}^{\text{xp}}$ with the MEP.
Note that by construction, any motif $M$ member of the training set will satisfy
$$
N_{e}^{\text{th}}(M)=N_{e}^{\text{xp}}(M)
$$
Thus, the compatibility with the MEP can only be evaluated on the test set.
However, this compatibility depends on the choice of the training set, since this choice determines the map $\gamma$.

\subsubsection{The MEP compatibility as a sampling artifact}
This makes difficult an error-based evaluation of the MEP for a given empirical ECTN distribution.

It would be easier to check a property that does not depend on the particular expression of the map $\gamma$.
An example of such a property is the following.
If $N_{e}$ satisfies the MEP, then we have
$$
\forall m,m', \forall M,M'\in E_{m,m'}\cup E_{m',m},N_{e}(M)=N_{e}(M')
$$

To check this fact in a given ECTN distribution $N_{e}$, we gather the observed ECTN into disjoint sets:
$$
S_{m,m'}=\left\{M\in E_{m,m'}\cup E_{m',m}|N_{e}(M)>0\right\}
$$
Two motifs belonging to the same set $S_{m,m'}$ should have the same probability.
Hence the ratio $\frac{\sigma_{m,m'}}{\mu_{m,m'}}$ should be small, where $\mu_{m,m'}$ is the average observed probability for the ECTN in $S_{m,m'}$:
$$
\mu_{m,m'}=\frac{1}{\left|S_{m,m'}\right|}\sum_{M\in S_{m,m'}}N_{e}(M)
$$
and $\sigma_{m,m'}$ is the dispersion of $N_{e}$ inside $S_{m,m'}$.

Sampling the values of $\frac{\sigma_{m,m'}}{\mu_{m,m'}}$ for every pair of NCTN $(m,m')$ allows to test in which extent two ECTN sharing the same sub-NCTN do have the same probability to be observed.
However, pairs $(m,m')$ for which $\left|S_{m,m'}\right|=1$ should be avoided since they lead systematically to $\sigma_{m,m'}=0$, which would constitute a bias in favour of the property we seek to test.

\begin{figure*}
    \subfigure[conf16]{
        \includegraphics[width=\columnwidth]{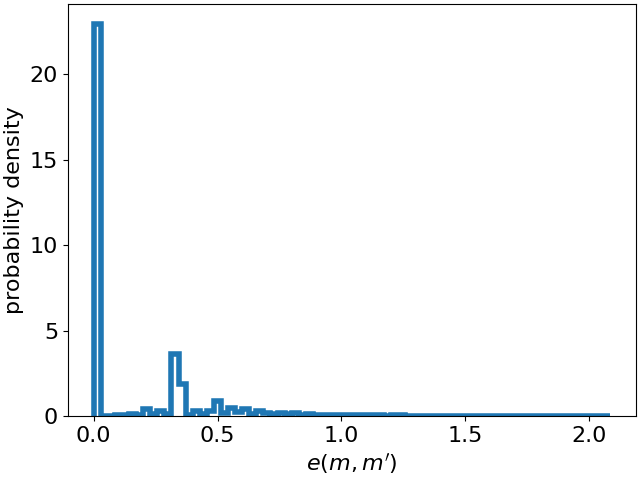}
    }
    \subfigure[utah]{
        \includegraphics[width=\columnwidth]{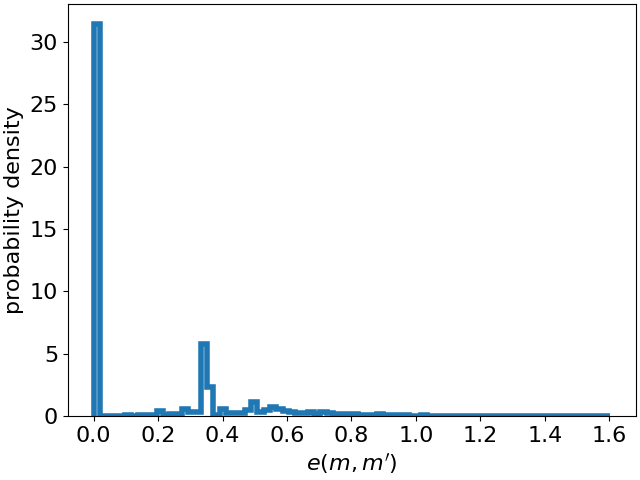}
    }
    \subfigure[work2]{
        \includegraphics[width=\columnwidth]{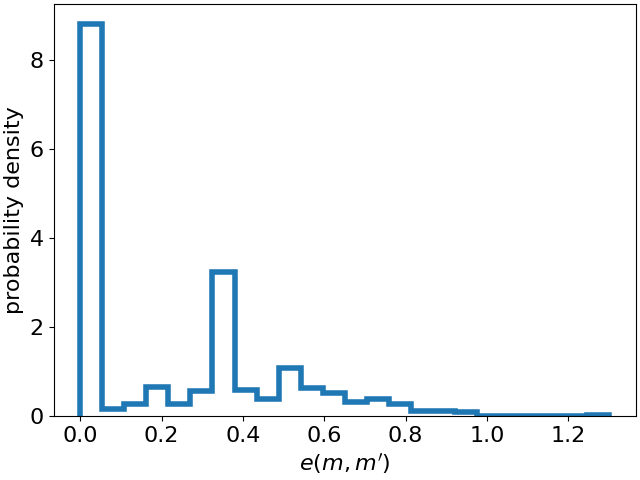}
    }
    \subfigure[ADM9conf16]{
        \includegraphics[width=\columnwidth]{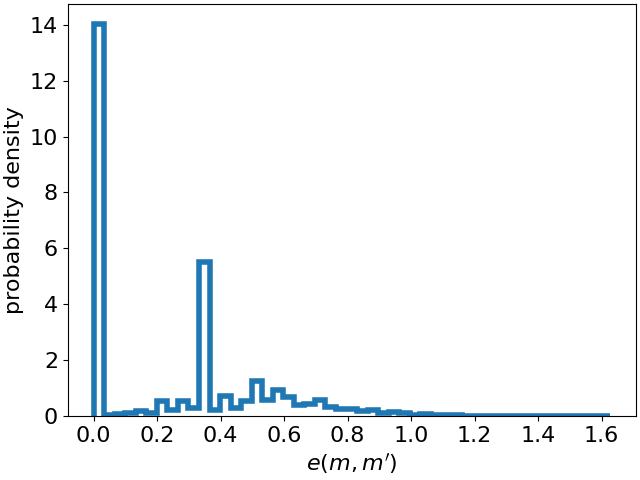}
    }
    \caption{\textbf{Checking the MEP property for some data sets.}
    (a) conference
    (b) primary school
    (c) workplace
    (d) a model
    ECTN of depth 3 are computed for each data set and the histogram of $\frac{\sigma}{\mu}$ is sampled for each data set.
    For a data set satisfying the MEP property perfectly, the histogram should be a Dirac located at zero.
    In practice, two modes are generally visible, the first one being close to zero -the compatible mode- and the other being wider -the outlier mode.
    The former is constituted by the ECTN that appear to be compatible with the MEP property, while the latter is constituted by outliers with respect to this property.
    }\label{fig:MEP_prop}
\end{figure*}

The histogram of $\frac{\sigma}{\mu}$ is shown on fig \ref{fig:MEP_prop} for some data sets and motifs of depth 3.

We see that for most data sets, this histogram is bimodal, with one mode satisfying the MEP property -the compatible mode- and one mode which does not -the outlier mode.
What characterizes the motifs belonging to either mode?

To answer this question, we separate the two modes and identify which ECTN contribute to each mode.
Then we compare how frequent are the ECTN contributing to the compatible mode with respect to the ECTN contributing to the outlier mode.

\begin{figure}
    \subfigure[conf16]{
        \includegraphics[width=0.46\columnwidth]{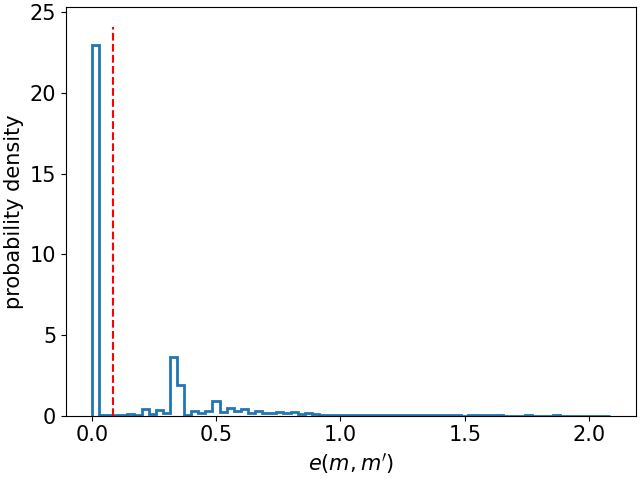}
    }
    \subfigure[highschool3]{
        \includegraphics[width=0.46\columnwidth]{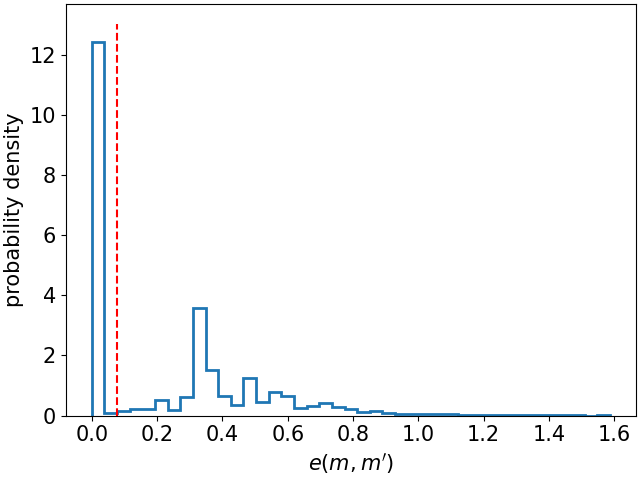}
    }
    \caption{\textbf{Automatic separation between the outlier and compatible modes.}
    Given an histogram of the MEP ratio (see text), we want to distinguish between the compatible mode, which gathers the ECTN that are compatible with the MEP, and the outlier mode, which gathers the ECTN which are not compatible with the MEP.
    The separating threshold is chosen as the local minimum between the two maxima corresponding to the two modes.
    Here it is indicated as a red dashed vertical line.
    The ECTN considered are a depth of 3 and the data sets are (a) a conference and (b) a highschool.
    }
    \label{fig:separation_modes}
\end{figure}

We localize the threshold $x_{th}$ separating the two modes according to the following procedure:
consider a distribution $\rho(x)$ for $x\geq0$, then $x_{th}$ is the largest point such that $\rho$ is decreasing on $[0,x_{th}]$.
The outcome of this procedure is illustrated on Figure \ref{fig:separation_modes} for two empirical data sets.

\begin{figure}
    \subfigure[conf16]{
        \includegraphics[width=0.46\columnwidth]{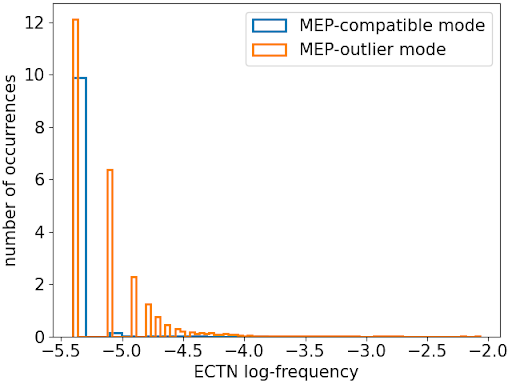}
    }
    \subfigure[highschool3]{
        \includegraphics[width=0.46\columnwidth]{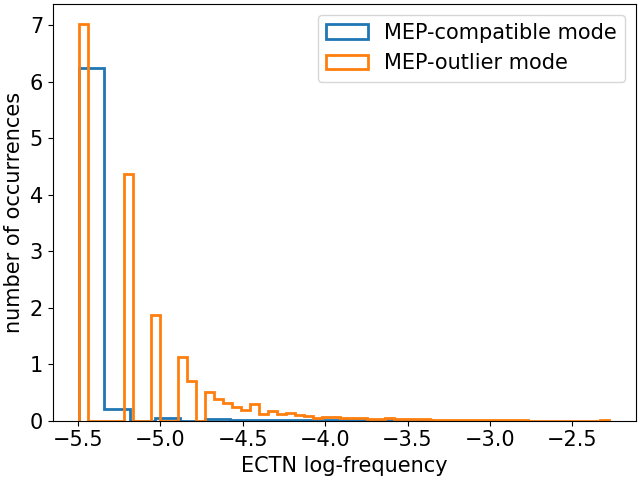}
    }
    \caption{\textbf{Frequency histogram of the ECTN belonging to each MEP mode.}
    For a conference (a) and a highschool (b), we sample separately the frequency histograms from the ECTN contributing to the compatible mode and the ECTN contributing to the outlier mode.
    We find that the MEP compatible mode is produced by very rare ECTN, while this is not the case of the outlier mode.
    This trend repeats for every other empirical data set we had at hand.
    This suggests that the MEP compatible mode may be due to a sampling artifact.
    }
    \label{fig:freq_modes}
\end{figure}

Once the ECTN contributing to each mode have been identified, we can now compare their properties.
As we announced, the first property we investigated is the ECTN frequency.
We displayed the frequency histogram on Figure \ref{fig:freq_modes} for two empirical data sets.

Clearly, we see that the compatible mode contains almost exclusively ECTN that occur a very few times in the whole data set, whereas the outlier mode contains also frequent ECTN.
Thus, the compatible mode may be a mere sampling artifact.
To check this, before computing the error ratio $e(m,m')=\frac{\sigma_{m,m'}}{\mu_{m,m'}}$, we remove the pairs of NCTN $(m,m')$ that produce very rare ECTN, i.e. such that $\forall M\in S_{m,m'},N_{e}(M)=1$.

\begin{figure}
    \subfigure[conf16]{
        \includegraphics[width=0.46\columnwidth]{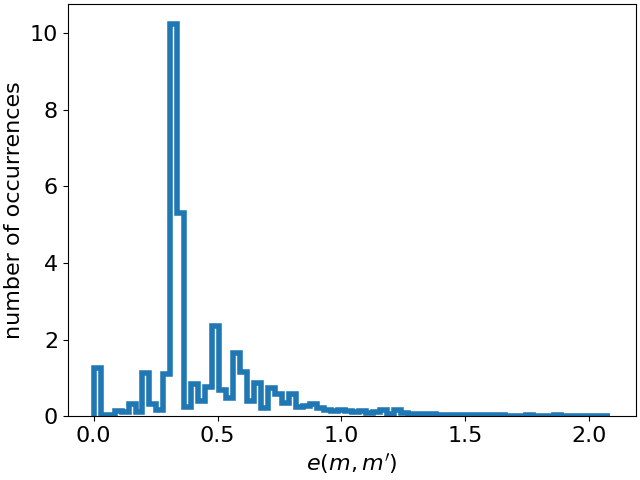}
    }
    \subfigure[highschool3]{
        \includegraphics[width=0.46\columnwidth]{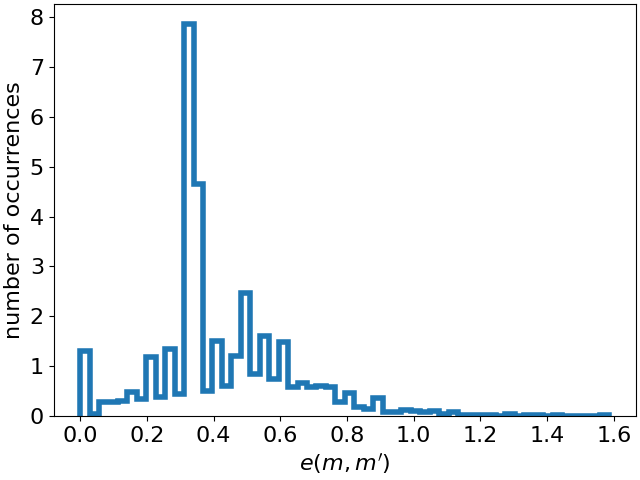}
    }
    \caption{\textbf{Filtered MEP error ratio histogram.}
    The MEP error ratio $e(m,m')$ is computed on unordered pairs of NCTN $(m,m')$.
    However, some precautions should be taken to avoid sampling artifacts.
    One precaution -described earlier- is to remove the pairs $(m,m')$ such that $S_{m,m'}$ contains a single ECTN, since $e(m,m')$ will trivially be zero.
    Another precaution is to remove the $(m,m')$ that contribute only to very rare ECTN:
    these are the $(m,m')$ such that $\forall M\in S_{m,m'},N_{e}(M)=1$.
    Applying these two precautions before sampling the error ratio leads to the filtered MEP error ratio histogram.
    We see that both in the case of a conference and a highschool (same data sets as in Figure \ref{fig:separation_modes} and Figure \ref{fig:freq_modes}), the MEP compatible mode does not survive to this filtering.
    This is actually the case for every empirical data set we tested, so empirical ECTN distributions are \textit{a priori} not compatible with the principle of maximum entropy.
    }
    \label{fig:filtering_modes}
\end{figure}

Actually, this filtering is enough to completely remove the compatible mode in every empirical data set, as we can see on Figure \ref{fig:filtering_modes} for two examples.

We deduce that empirical ECTN distributions are \textit{not} compatible with the principle of maximum entropy.
What does this tell us about the empirical data?
If the MEP were satisfied by the empirical ECTN distribution, this would mean that the ECTN distribution does not contain any additional information with respect to the NCTN distribution.
In this eventuality, sampling ECTN besides NCTN would be superfluous.
Here, we showed that recovering the ECTN distribution would require more constraints than its compatibility with the NCTN distribution.
This indicates that the knowledge of ECTN in empirical networks does bring some information about this network that is not included in the NCTN distribution.
It remains to determine the nature of this information, as will be attempted in the next section.

\subsubsection{How far are empirical data sets from the MEP}

\begin{figure}
    \subfigure[Computation of the coordinates]{
        \includegraphics[width=\columnwidth]{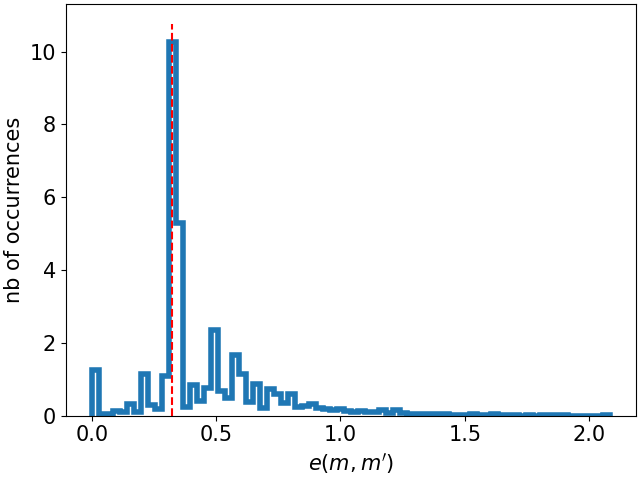}
    }
    \subfigure[MEP compatibility in logarithmic scale]{
        \includegraphics[width=\columnwidth]{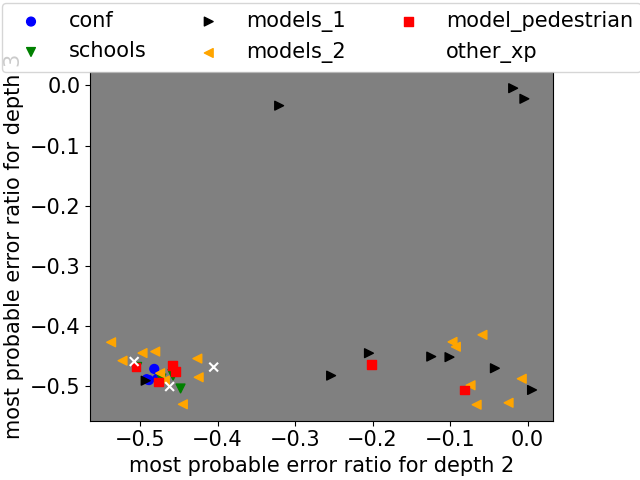}
    }
    \caption{\textbf{Quantifying how far is a temporal network from MEP compatibility.}
    (a) The red dashed vertical line indicates the most probable error ratio \textit{after} the filtering described in Figure \ref{fig:filtering_modes}.
    The lowest this ratio, the most MEP compatible is the ECTN distribution.
    (b)
    Computed for depths 2 and 3, this ratio serves placing temporal networks as points in a 2D plane.
    Here, the axes have a logarithmic scale.
    We see that empirical data sets share similar compatibility, and that this is the highest compatibility among the data sets tested here.
    }
    \label{fig:how_far_MEP}
\end{figure}

Although we have shown that the principle of maximum entropy does not account for the empirical ECTN distributions, we may ask how incompatible are empirical data sets with the principle of maximum entropy.
To answer this, one way is to characterize a temporal network by a pair of coordinates $(C_{2},C_{3})$ in the spirit of Figures \ref{fig:compa_hyp1} and \ref{fig:MEP_hyp_compa}, where $C_{d}$ denotes the most probable value of the filtered MEP error ratio for ECTN of depth $d$.
The procedure and result are displayed on Figure \ref{fig:how_far_MEP}.

Surprisingly, the most probable error ratio seems to be shared across empirical data sets and is the same for depths 2 and 3.
Also, we see that empirical data sets are more compatible with the MEP than models:
the most compatible models are just as compatible as the empirical data sets.

Future investigation would be needed to understand why it is so, and whether the principle of maximum entropy would finally be a useful tool to characterize the empirical ECTN distributions.
Another direction of research would be to understand what kind of graph-valued stochastic processes would satisfy the principle of maximum entropy.

\subsection{Testing the hypothesis of independent edges (hyp 1)}\label{subsec:ind_edges}

Now we have exlcuded the MEP hypothesis, let us test whether the ECTN distribution in empirical networks could be derived from assuming independent edges.
In the (highly probable) case of a negative answer, we want to identify the reason why.

\begin{figure*}
    \subfigure[linear scale]{
        \includegraphics[width=\columnwidth]{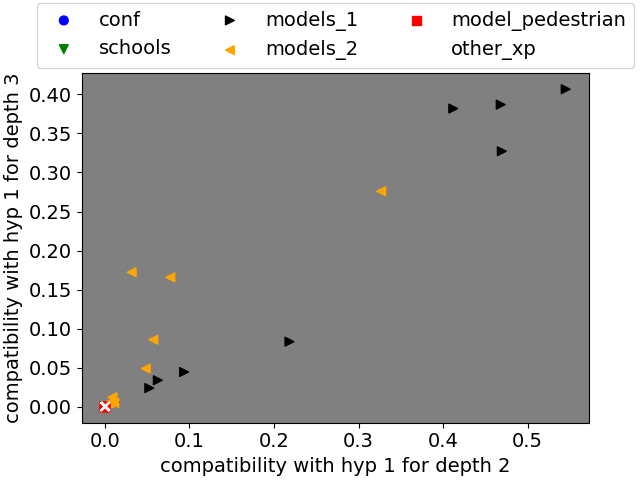}
    }
    \subfigure[logarithmic scale]{
        \includegraphics[width=\columnwidth]{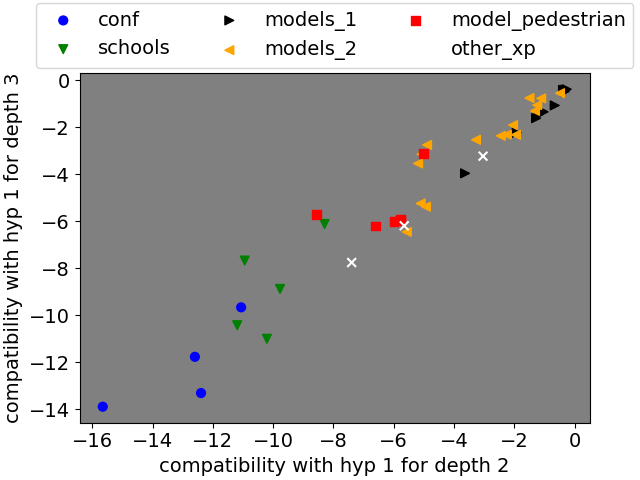}
    }
    \caption{\label{fig:compa_hyp1}\textbf{Compatibility of various data sets with the hypothesis of independent edges.}
    (a) linear scale: we see that the data sets for which the hypothesis suits best are all models.
    (b) logarithmic scale: the coordinates are now $(\log_{10}(C_{2}),\log_{10}(C_{3}))$.
    This allows to have a better resolution in the low compatibility region.
    In particular, we see that the empirical data sets are the less explained by the hypothesis, with the worst compatibility being as low as $10^{-14}$.
    Also, notice that all the data sets are close to the diagonal, meaning that the compatibility does not vary much between depths 2 and 3.
    }
\end{figure*}

Following the same procedure as in the previous subsection, we compute the coordinates of our data sets in the compatibility plane $(C_{2},C_{3})$ for the hypothesis of independent edges.
Results are drawn on Figure \ref{fig:compa_hyp1}.

This figure shows that the less compatible data sets are empirical data sets.
Some models are well explained by our hypothesis, in particular the model ``min\_EW3'', which has strictly zero spatial correlations and thus should display a compatibility of 1.
However, it is a bit lower in practice, due to statistical fluctuations.
For the measure we use, fluctuations are significant and always decrease the compatibility.

\begin{figure*}
    \subfigure[depth 2]{
        \includegraphics[width=\columnwidth]{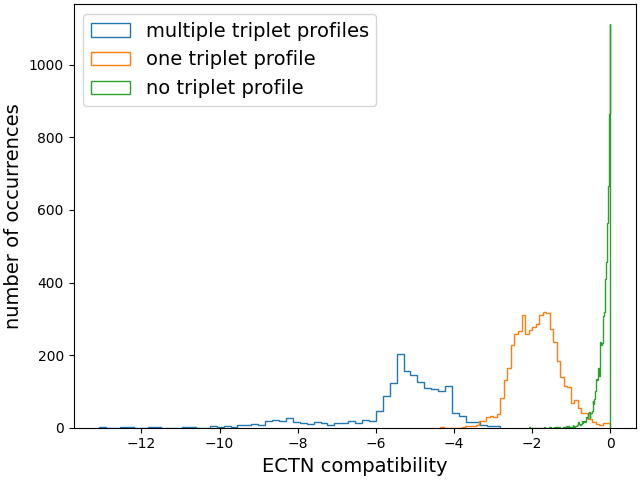}
    }
    \subfigure[depth 3]{
        \includegraphics[width=\columnwidth]{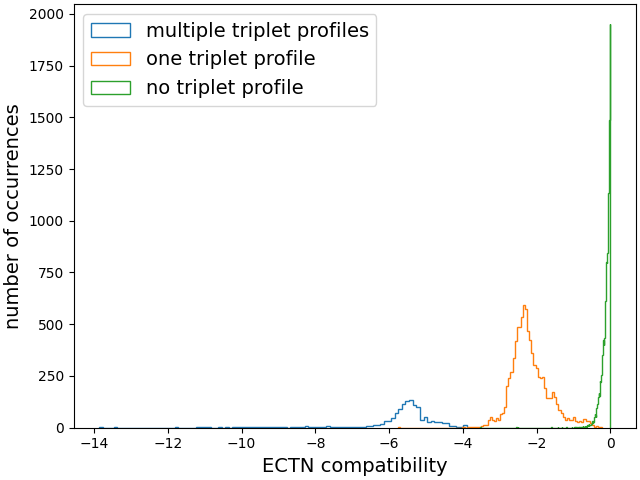}
    }
    \caption{\label{fig:low_compa}\textbf{Identifying the motifs of lower compatibility.}
    Contrary to Figure \ref{fig:low_compa_MEP}, the distributions are multimodal.
    Each mode is constituted by the motifs containing the same number of triplets (cf text for definition).
    Intuitively, this means that assuming independent edges yields fewer triangles than observed in empirical data sets.
    (a) ECTN of depth 2.
    (b) ECTN of depth 3.
    }
\end{figure*}

Overall, assuming independent edges is not compatible with the ECTN distribution in empirical temporal networks.
Now this is clear, can we identify the reason why?
To do this, we display the compatibility histogram on Figure \ref{fig:low_compa}, that we compute similarly to Figure \ref{fig:low_compa_MEP}.

These distributions are multimodal, one mode being of high compatibility and the others being of low compatibility.
By tracking which motifs contribute to each mode, we can establish that the difference between motifs from distinct modes is purely structural.
It means that the mode to which a motif contributes can be deduced from its expression (e.g. string or map representation) solely.

By numbering the modes $c=0,1,\hdots$ by decreasing compatibility, it turns out that the mode $c$ is constituted by the motifs containing exactly $c$ triplets, where we will define this notion now.

In intuitive terms, a triplet is a node which interacts with at least two distinct neighbours within $d$ consecutive time steps.
In terms of interactions, it corresponds to a node whose degree is equal or higher than 2 when aggregated on $d$ consecutive time steps.
Considering the string representation of motifs, a triplet is an ECTN satellite profile containing either at least one letter `3' or at least one letter `1' and one letter `2'.

The number of triplets in a given ECTN is the number of satellites which are triplets.
Although the compatibility of an ECTN becomes low as soon as it contains one triplet, the ECTN without any triplets are highly compatible.
This indicates that, even in empirical data sets, adjacent edges are not always correlated with each other.
In particular, different triplets might not be correlated, while the edges belonging to the same triplet clearly are.
Since different triplets are associated to different nodes, we should now investigate the hypothesis of independent nodes, which we have promoted to the hypothesis of conditional independence of edges.

\begin{figure*}
    \subfigure[linear scale]{
        \includegraphics[width=\columnwidth]{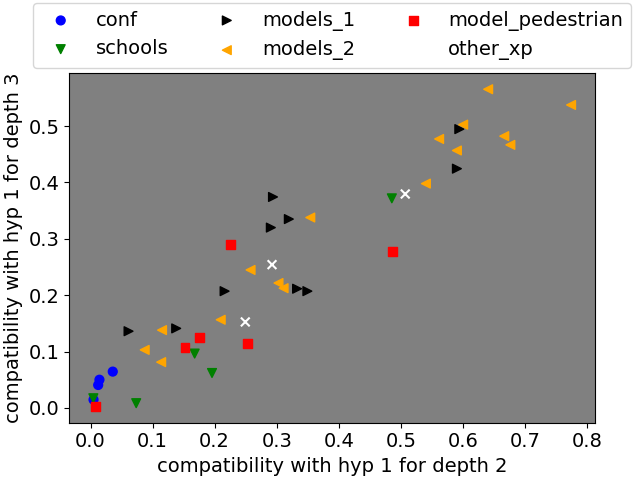}
    }
    \subfigure[logarithmic scale]{
        \includegraphics[width=\columnwidth]{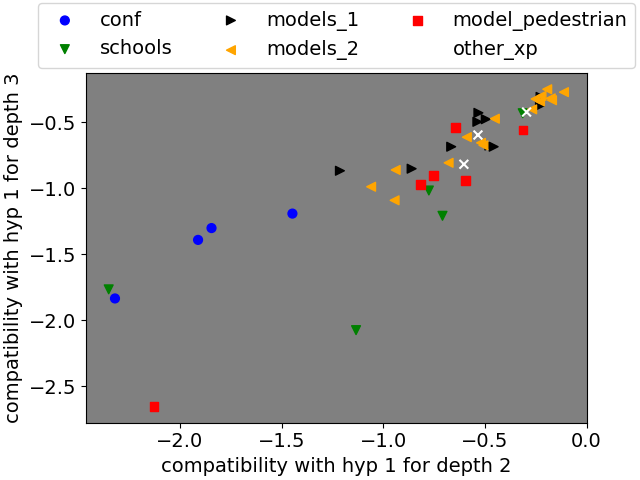}
    }
    \caption{\textbf{Compatibility of the NCTN distribution with the hypothesis of independent edges.}
    This Figure reads the same way as Figure \ref{fig:compa_hyp1}, but with NCTN instead of ECTN.
    (a) linear scale:
    in contrast to Figure \ref{fig:compa_hyp1}, 5 empirical data sets have a compatibility above 0.1 for both depths.
    (b) logarithmic scale:
    Like in the case of ECTN, the conferences are the empirical data sets with the lowest compatibility.
    However, this compatibility has increased by 12 orders of magnitude as compared to the case of ECTN.
    }\label{fig:compa_NCTN}
\end{figure*}

However, before doing that, let us mention how does the hypothesis of independent edges compare with the NCTN distribution.
We use the same procedure as just described, the results being visible on Figure \ref{fig:compa_NCTN}.

We see that assuming independent edges is much more compatible with the NCTN than the ECTN distribution.
There is also much diversity across the empirical data sets:
the most dense data sets (conferences) are the less compatible while the sparsest data sets (workplace, hospital) are the most compatible.
Said otherwise, it seems that when interactions are rare, they become independent from each other from the point of view of NCTN; this may indicate that in a low-activity regime, activity profiles of edges are weakly interacting, probably because of a smaller clustering coefficient.
This low clustering coefficient comes with a low number of triplets, which is precisely the cause of failure for the hypothesis of independent edges in the case of ECTN.
However, future investigation is required to understand whether the cause of failure in the case of NCTN is the same as for ECTN.
Also we should try to understand why assuming independent edges seems to describe well many NCTN empirical distributions, while it is clearly not the case for ECTN.

\subsection{Testing the conditional independence of edges (hyp 2)}

To test whether we can obtain the empirical ECTN distributions when considering that edges are independent conditioned to some hidden variables, we should minimize the cost function from equation \ref{eq:cond_ind} in sub-subsection \ref{subsubsec:cond_ind_test}.
This minimization should be performed under the constraints precised in the same sub-subsection.
As it cannot be done analytically, we used the numerical solver scipy.optimize.least\_squares to get an approximate solution.
Doing so requires specifying the number of hidden variables, which we set to 2 in our numerical investigations.

\begin{figure*}
    \subfigure[Linear scale]{
        \includegraphics[width=\columnwidth]{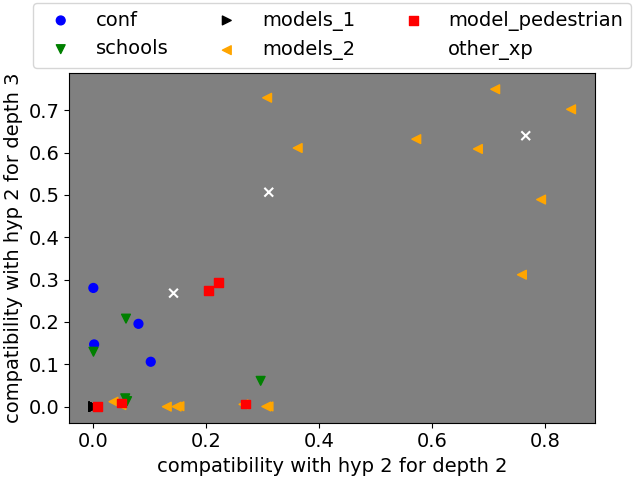}
    }
    \subfigure[Logarithmic scale]{
        \includegraphics[width=\columnwidth]{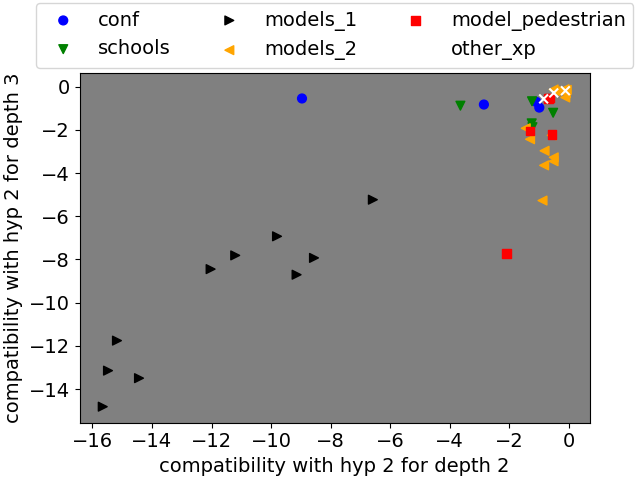}
    }
    \caption{\textbf{Compatibility of various data sets with the hypothesis of conditional independent edges.}
    As in Figure \ref{fig:compa_hyp1}, the data sets are scattered through the 2D plane, both in linear (a) and logarithmic scales (b).
    In particular, we see that four empirical data sets have a compatibility above 0.1 for both depths 2 and 3.
    Also, only two empirical data sets (conferences) have a compatibility below 0.01 for either depth.
    }
    \label{fig:cond_ind_comp}
\end{figure*}

Once the ECTN theoretical distribution $P^{\text{th}}$ has been computed, we evaluate how close it is to the empirical distribution $P^{\text{xp}}$ by using the same definition for compatibility as subsection \ref{subsec:ind_edges}.
The results are shown on Figure \ref{fig:cond_ind_comp}.
Clearly, the ECTN empirical distributions seem much more compatible with conditioned independence than independence alone.
Comparing the panels (b) of Figures \ref{fig:cond_ind_comp} and \ref{fig:compa_hyp1}, we notice that the ADM models have much decreased in compatibility.
This is surprising because the conditioned independence is a weaker hypothesis:
the compatibility with it should always be higher than the compatibility with the hypothesis of independent edges.

However, we do not compute $P^{\text{th}}$ the same way in both cases:
it is analytically derived for the hypothesis of independent edges, while for the hypothesis of conditioned independence we minimize a cost function.
Given this function is not convex, it may have many local minima, that make difficult to find the global minimum in some cases.
Additional research has to be done in order to assess this explanation, and to determine an efficient minimization procedure that allows to find the global minimum.
For example, stochastic gradient descent may be useful, as inspired from the case of deep learning \cite{amari1993backpropagation}:
instead of computing the error averaged over all ECTN, it may be more efficient to pick one ECTN $M$, minimize the one sample error $\left(P^{\text{th}}(M;f,\Omega,p)-P^{\text{xp}}(M)\right)^{2}$, then pick another ECTN and repeat until convergence of the parameters $f,\Omega,p$.

To conclude this sub-section, the hypothesis of conditional independence seems to be a promising explanation for the empirical ECTN distributions, which should be investigated further in future research. 

\section{the graph tiling theory: how to infer the network topology from the ECTN distribution}

\subsection{About gluing motifs to build networks}

In this paper, we tried to identify what kind of information were conveyed by node and edge-centered motifs.
In particular, we showed that motifs could be used to probe for spatial correlations in temporal networks.
However, we only considered the total number of occurrences of motifs, disregarding the fact that they are not distributed at random among a temporal network:
to give a simple example, when the number of interactions increases, motifs that end with a greater number of interactions will be more abundant.
Besides being temporally distributed, motifs are likely to be spatially distributed, some motifs being neighbours of others in a not-random way.

This problem is the gluing problem, that we propose under this form:
\begin{itemize}
    \item How to glue ECTN motifs with each other to recover the original network?
    \item How many networks are compatible with a given ECTN distribution?
    \item What kind of network do we obtain if we glue at random the ECTN motifs from a given distribution?
\end{itemize}

The first question is studied in the work of Longa and al. \cite{longa2024generating}.
Here, we will show some preliminary work that attempts to answer the second question.
The third question is postponed to future work.

\subsubsection{Tree motifs}

As we want to start as simple as possible, we will consider static connected networks and ECTN of depth 1.
In particular, we will pay attention to the networks whose ECTN distribution contains a very few motifs.
On the contrary, empirical networks contain typically hundreds or thousands of different ECTN.

The generic question is the following:
given an ECTN distribution, how many simple connected networks of infinite size can be built that are compatible with that distribution?

To investigate this question, we will consider distributions that contain an increasing number of ECTN:
first we will consider Dirac distributions, then distributions with two motifs, etc.
We will also see what can happen when we impose the network to be of finite size, which is the case of empirical networks.

To allow the reader to familiarize with this question, we will start by simple examples.
Consider the following ECTN distribution:
$$
P_{e}(M)=\delta_{M,12}
$$
where we have used the string representation.
Since we consider motifs of depth 1, the central edge has only one possible profile $c=1$ so we omit it in the ECTN string.
Thus, 12 corresponds to the ECTN isomorphic to a line graph with four nodes.

The only infinite simple connected graph (in the remainder of this section, ``graph'' will actually stand for ``simple connected graph'') compatible with this Dirac distribution is the infinite line graph.
If we truncate this graph so that it becomes finite, two possibilities emerge:
One possibility is to glue the two ends of the line graph with each other, in which case we obtain a cycle (a circle graph), which has the same ECTN distribution.
The other possibility is to let the two ends naked, in which case we obtain a finite line graph.
For a graph with $N$ nodes, this results in the following ECTN distribution:
$$
P'_{e}(M)=\frac{1}{N-1}\left[
(N-3)\delta_{M,12}+2\delta_{M,1}
\right]
$$
To compare this situation with the empirical case, we should take into account that a snapshot typically contains many connected components, with a large distribution for the number of nodes $N$.
Having such a diversity in size of the connected components will mix together multiple ECTN distributions, but here, in this preliminary work, we restrict ourselves to a single connected component.
The mixing of distinct distributions should be the focus of future work.

With this simple example, we see that (1) having a Dirac ECTN distribution constraints enormously the compatible networks and (2) finite size effects can alter the ECTN distribution in a predictable way.
Some motifs are not compatible with any infinite graph:
e.g. no infinite graph is compatible with the ECTN 1.
We can actually prove the following result:
A Dirac ECTN distribution $P_{e}(M)=\delta_{M,M_{0}}$ with $M_{0}$ containing no letter 3, is compatible with at least one infinite graph if and only if $M_{0}$ contains \textit{both} at least one letter 1 and one letter 2.
When so, the number of infinite compatible graphs is generally infinite.
In general, there is also an infinite number of compatible graphs of finite size.

\begin{figure}
    \centering
    \subfigure[most regular construction]{
        \includegraphics[width=\columnwidth]{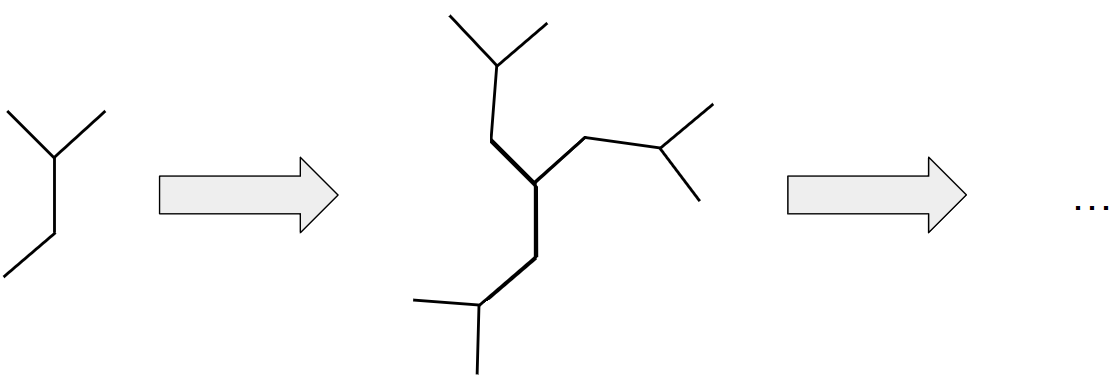}
    }
    \subfigure[construction with a cycle]{
        \includegraphics[width=\columnwidth]{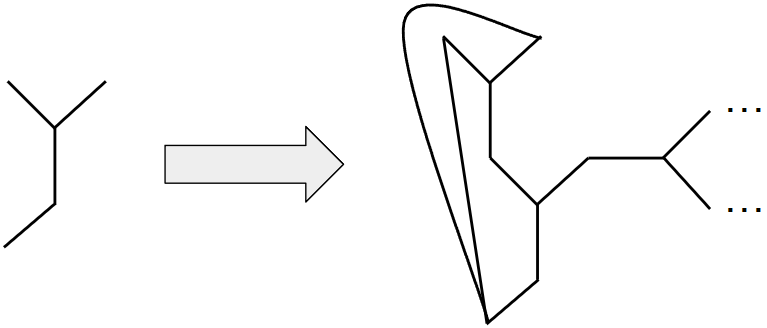}
    }
    \caption{\textbf{Construction of an infinite network compatible with a Dirac ECTN distribution.}
    Here we consider infinite graphs containing only 112 as ECTN.
    There are a infinite number of such graphs, differing by their cycles.
    (a) Most regular construction of a compatible graph:
    it contains no cycle.
    (b) Example of an infinite compatible graph which has the smallest possible cycle (4 nodes) compatible with the considered ECTN.
    In any case, the degree sequence of any node will be periodic of period 2, and determined by the motif contained in the graph.
    }
    \label{fig:dirac_motif}
\end{figure}

An example of the construction of a compatible graph is given on Figure \ref{fig:dirac_motif}.

These compatible graphs correspond exactly to the graphs with an alternating degree sequence $K=[K_{1},\hdots]$ of period 2, with $K_{n}$ being the degree of the $n^{\text{th}}$ order neighbour.
If we start from some node, the degree sequence will look like this:
$$
K=[n_{1}+1,n_{2}+1,\hdots]
$$
where $n_{1}$ (resp. $n_{2}$) is the number of letters 1 (resp. 2) inside $M_{0}$.
If we choose another starting node, the sequence may start with $n_{2}+1$ instead.

\subsubsection{Motifs with triangles}

\begin{figure}
    \centering
    \includegraphics[width=\columnwidth]{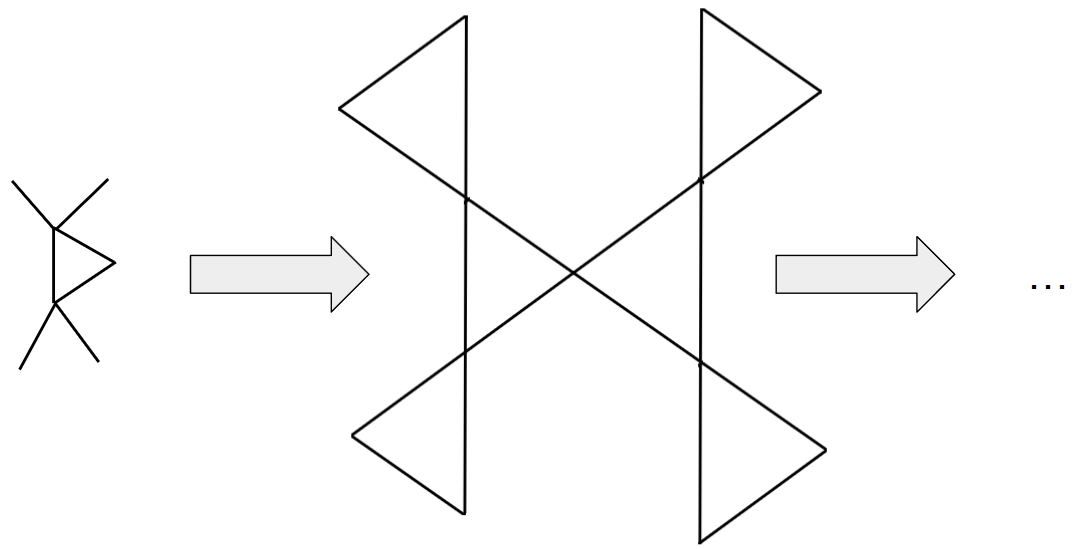}
    \caption{\textbf{Construction of an infinite network compatible with an ECTN containing a triangle.}
    Here we consider the case of the ECTN 11223, which can grow an infinite network.
    }
    \label{fig:dirac_triangle}
\end{figure}

What happens when we consider motifs that contain at least one letter 3, i.e. $n_{3}\neq0$?
The simplest such motif writes 3, and is not compatible with any infinite graph.
We can check that 123 is in the same case as well.
Hence, now that $n_{3}\neq0$, it is not enough to have $n_{1},n_{2}\neq0$ to allow infinite graphs to grow.
To find one motif that can, it is easier to draw an infinite regular graph and compute the associated motifs.
Doing so, we find that the ECTN 11223 can grow an infinite graph, which is shown in Figure \ref{fig:dirac_triangle}.

More generally, what are the motifs with $n_{3}\neq0$ that can grow an infinite graph?
First, let us show that they satisfy:
$$n_{1}=n_{2}\neq0$$
To do this, let us assume that the motif $\vec{n}=(n_{1},n_{2},n_{3})$ with $n_{1}\neq n_{2}$ can grow an infinite graph.
In such a graph, any node $i$ has two possible values for its degree:
$$
k_{i}\in\{n_{1}+n_{3}+1,n_{2}+n_{3}+1\}
$$
Moreover, if $i$ and $j$ are adjacent to each other, then $k_{i}\neq k_{j}$ since $n_{1}\neq n_{2}$.
Then we get a contradiction by considering a node $l$ such that $ijl$ is a triangle ($l$ exists because $n_{3}\neq0$):
the degrees of $i,j,l$ must be all distinct but only two values are available.
Thus we must have $n_{1}=n_{2}$.
On the other hand, the motif 3 cannot grow an infinite graph so we must have $n_{1}\neq0$.
Note that the same line of reasoning allows us to establish that an infinite graph generated by a single motif with $n_{3}=0$ and $n_{1}\neq n_{2}$ can only contain cycles involving an even number of nodes:
this is a property of any graph endowed with an alternating degree sequence of period 2.

Now we have shown that $n_{1}=n_{2}$, let us denote a motif containing triangles by $(n,n_{3})$, where $n_{1}=n_{2}=n$.
What are the values of $n$ and $n_{3}$ that allow to grow an infinite graph?

For $n_{3}=1$, we can answer the question.
In this case, triangles in the infinite graph cannot share a common edge (otherwise this edge will satisfy $n_{3}\geq2$).
The only possibility to grow an infinite graph is to glue the triangles at the level of the nodes, like in the example of Figure \ref{fig:dirac_triangle}.
Thus, if $n_{3}=1$, the possible values for $n$ are the $n=2p$, with $p\in\mathbb{N}^{*}$.
The resulting graphs will only contain cycles of length 3, and for a given value of $p\geq1$, there is a single infinite graph which is compatible with the motif $(2p,1)$.

\begin{figure}
    \centering
    \includegraphics[width=\columnwidth]{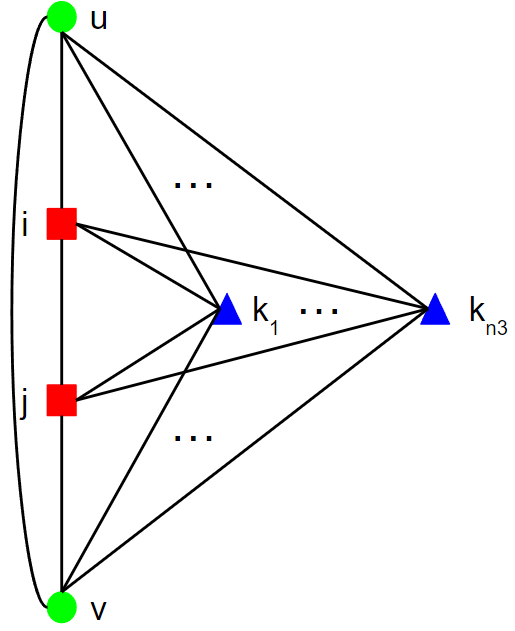}
    \caption{\textbf{Proving that $n=1$ cannot yield an infinite graph.}
    If we try to draw an infinite graph that contains only the ECTN $(n=1,n_{3}\geq2)$, we inevitably encounter the graph drawn here.
    The first edge to be drawn is the edge $(i,j)$, with $i$ and $j$ represented as red squares.
    The green circle nodes $u$ (resp. $v$) correspond to the upper (resp. lower) stubs of the edge $(i,j)$.
    The blue triangle nodes $k_{1},\cdots,k_{n_{3}}$ are the neighbours common to $i$ and $j$.
    When we try to grow an infinite graph starting from $(i,j)$, it turns out (cf main text) that each $k_{n}$ must link to every other except one node.
    This enforces every node of the graph to reach its final degree of $n_{3}+2$, which prevents the growth to pursue any further.
    }
    \label{fig:n1_graph}
\end{figure}

For $n_{3}\geq2$, we can show that we must have $n\geq2$.
Indeed, $n=0$ does not work so let us try to grow an infinite graph with $n=1$.
We begin by drawing the motif $(1,n_{3})$ and we name the central edge $(i,j)$ and $k_{1},\hdots,k_{n_{3}}$ the neighbours linked both to $i$ and $j$.
We name $u$ the last neighbour of $i$ and $v$ the last neighbour of $j$ (cf Figure \ref{fig:n1_graph}).
Then, the only possibility for the edge $(i,u)$ to have $n_{3}$ triangles and only a single upper and lower stubs is to link to each of the $k_{n}$.
Otherwise, $(i,u)$ would have at least two lower stubs, e.g. the edges $(i,j)$ and $(i,k_{1})$.
As $j$ cannot connect to $u$, $k_{1}$ must, resulting in the triangle $(i,u,k_{1})$.
The story is similar for the edge $(v,j)$.
Let us now consider the edge $(k_{1},u)$.
At this stage, it has two lower stubs $(k_{1},j)$ and $(k_{1},v)$.
As $j$ cannot connect to $u$, the edge $(u,v)$ must exist, resulting in the triangle $(u,k_{1},v)$.
Then, $(k_{1},u)$ has the $n_{3}-1$ upper stubs $(u,k_{2}),\cdots,(u,k_{n_{3}})$.
All these stubs must be absorbed in triangles except one, which implies that $k_{1}$ must link to each but one of the $(k_{n})_{n\neq1}$.
Note that these links are not drawn on Figure \ref{fig:n1_graph}.
As the story is similar for every $n$ between 1 and $n_{3}$, we deduce that we have grown a graph with every node having a degree of $n_{3}+2$.
As this degree equals the degree any node should have in the final infinite graph, we deduce that our finite graph cannot grow any further, since any addition of an edge will result in at least one node having a degree greater than the maximal allowed degree.
Hence, we must have $n\geq2$.

To get further insight, we can rewrite our problem in terms of the adjacency matrix.
Finding a connected graph that contains $(n,n_{3})$ as unique ECTN is equivalent to find a square symmetric matrix $A$ such that
$$
\begin{cases}
    A_{i,j}\in\{0,1\}\land A_{i,i}=0
    \\
    A\text{ corresponds to a connected graph}
    \\
    AU=(n+n_{3}+1)U
    \\
    A_{i,j}=1\implies\left(A^{2}\right)_{i,j}=n_{3}
\end{cases}
$$
where $U$ denotes the vector containing only 1 as entries.
The first condition is due to the fact that $A$ is the adjacency matrix of an unweighted graph with no self-loops allowed:
$A_{i,j}=1$ if $i$ and $j$ are linked to each other and 0 otherwise.
The second condition translates the fact that every node has the same degree, equal to $n+n_{3}+1$.
The third condition translates the fact that every edge is involved in exactly $n_{3}$ triangles.

Then, the question to address is which values of $n$ and $n_{3}$ allow to find a solution $A$ when the size of $A$ is infinite.

From the constraints of $A$, we can deduce the number of triangles $\Delta(i)$ shared by a common node $i$.
We have on the one hand:
$$
(A^{3})_{i,i}=2\Delta(i)
$$
On the other hand:
\begin{align*}
    (A^{3})_{i,i}
    &=
    \sum_{k,l}A_{i,k}A_{k,l}A_{l,i}
    \\
    &=
    \sum_{l,A_{i,l}=1}\sum_{k}A_{i,k}A_{k,l}
    =\sum_{l,A_{i,l}=1}(A^{2})_{i,l}
    \\
    &=
    \sum_{l,A_{i,l}=1}n_{3}=n_{3}(n+n_{3}+1)
\end{align*}
We deduce that
$$
\Delta(i)=\frac{1}{2}n_{3}\left(n+n_{3}+1\right)
$$

As $\frac{1}{2}n_{3}\left(n_{3}+1\right)$ and $\Delta(i)$ are both integers, we deduce that $\frac{1}{2}n_{3}n$ is also an integer, i.e. $nn_{3}$ must be even.

\begin{figure}
    \centering
    \includegraphics[width=\columnwidth]{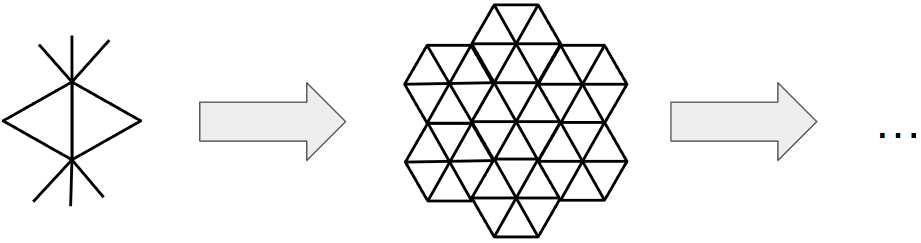}
    \caption{\textbf{Growing an infinite graph from the ECTN $(3,2)$.}
    To grow an infinite graph from the ECTN $(n=3,n_{3}=2)$, one possibility is first to tile the plane with hexagons.
    Then, add a node at the center of each hexagon, and link it to every vertex of the hexagon.
    Another possibility (not shown) is to tile the plane with squares and draw a single diagonal for each square.
    The diagonals should all be parallel to each other.
    This yields a graph that is actually isomorphic to the one drawn here.
    }
    \label{fig:n3_graph}
\end{figure}

Future research is required to determine what are all the ECTN $(n,n_{3})$ which are compatible with an infinite graph.
We show on Figure \ref{fig:n3_graph} an example of infinite graph with $n_{3}=2$, proving that the ECTN containing no or a single triangle are not the only ones that can grow an infinite graph.
To close this sub-subsection, let us recap the properties we have proven concerning $n$ and $n_{3}$ for the infinite graphs that contain a single ECTN $(n,n_{3})$ with $n_{3}\geq1$:
\begin{itemize}
    \item if $n_{3}=1$, then $n=2p$ with $p\in\mathbb{N}^{*}$
    \item for any $n_{3}\geq1$, we have that $n\geq2$ and $nn_{3}$ must be even
\end{itemize}

To get further constraints on $n$ and $n_{3}$, one possibility would be to bound the number of cycles of length 4.
Indeed, we have:
$$
\begin{cases}
    \text{Tr}(A^{2})=Nk
    \\
    \text{Tr}(A^{3})=Nkn_{3}=\left<A^{2}|A\right>\leq\sqrt{\text{Tr}(A^{4})}\sqrt{\text{Tr}(A^{2})}
\end{cases}
$$
where we used the inequality of Cauchy-Schwarz for the canonical scalar product on the space of square matrices:
$$
\left<A|B\right>=\text{Tr}(A^{T}B)
$$
We also introduced $N$ the number of nodes in the graph corresponding to $A$ and $k=n+n_{3}+1$ the degree of any node in that graph.

The last inequality can be rewritten as a constraint on $k$ and $n_{3}$:
$$
n_{3}^{2}\leq\frac{1}{Nk}\text{Tr}(A^{4})
$$

One rough upper bound on the diagonal coefficients $(A^{4})_{i,i}$ can be estimated:
$$
\frac{1}{Nk}\text{Tr}(A^{4})\leq
k+(k-1)^{2}-n_{3}
$$
Although this bound does not give any information about $n_{3}$ or $n$, it is not possible to grow an infinite graph for which this bound is reached.
Hence, it should be possible to find a lower upper bound, which may result in useful constraints on $n$ and $n_{3}$.
Finding the lowest upper bound is equivalent to determine the maximum possible number of cycles of length 4 in an infinite graph grown by a single ECTN.

\subsection{Distributions with multiple motifs and future directions of research}\label{subsec:tiling_th}

We have characterized some of the infinite graphs generated by a single motif.
However, empirical networks typically contain many different motifs.
Thus an interesting question would be:
given a family of size $p<\infty$ of ECTN $F=\{M_{1},\hdots,M_{p}\}$, can we build an infinite graph that contains all and only the elements of $F$ as ECTN?
Referring to such a family that is able to grow an infinite graph as a $p$-family, we ask the following questions:
\begin{itemize}
    \item what are the \textit{minimal} $p$-families?
    where a minimal $p$-family $F$ satisfies that removing any element from it will not yield a $(p-1)$-family.
    Said otherwise, $F$ is such that $\forall M\in F$, there does not exist any infinite graph whose ECTN distribution matches the family $F\setminus\{M\}$.
    \item what are the \textit{maximal} $p$-families?
    where a maximal $p$-family satisfies that adding any ECTN to it will not yield a $(p+1)$-family.
\end{itemize}
We gather those questions under the hood of ``graph tiling theory'', and propose them to the network science community.
Indeed, in the long run formal research in this direction could lead to insights about the structural complexity and properties of temporal networks of social face-to-face interactions.

To help such an investigation, we propose a concept, thought for exploratory analysis, which captures partly the way ECTN of a given depth are glued to each other in a temporal network.

Let us consider a temporal graph $G$ and let us restrict to ECTN of depth $d$, where $d$ can denote any non-zero integer.
We associate to $G$ and $d$ a weighted undirected hypergraph, called the \textit{ECTN gluing hypergraph}.
Denoting by $w_{I}$ the weight of the undirected hyperedge $I$, we define it as follows:
\begin{align*}
    w_{M_{1},\hdots,M_{p}}
    =&\text{number of instances of the ECTN }M_{1},\hdots,M_{p}
    \\
    ~~&\text{that all share at least one temporal edge}
\end{align*}
To build this graph, we can look at every temporal edge and identify the instances of ECTN in which it appears.
Then an hyperedge is drawn between those ECTN.
Note that a temporal edge can appear in different instances of the same ECTN, so the labels $I$ of the hyperedges are not sets:
as sets they are unordered, but unlike sets, they allow for repetitions of the same element.

\section{Conclusion}
In this paper, we have introduced the so called edge-centered temporal neighborhoods (ECTN) motifs, which can be seen as a generalization of the egocentric temporal neighborhoods (ETN) from the literature.
We have first shown how to recover the ETN distribution from the ECTN distribution, and checked that in empirical networks of face-to-face interactions, the ECTN distribution brings additional information with respect to the NCTN (ETN) motifs.
Then, we proposed different assumptions on the dynamics generating a network, investigating whether one of these assumptions could account for the observed ECTN distribution.
We found that the NCTN distribution was well approximated by the hypothesis of independent edges, while the ECTN distribution was better approximated by the hypothesis of conditional independence.
The last part of the paper is more the inauguration of a new mathematical framework than actual results:
we investigate how the topology of a network is constrained from its ECTN distribution alone.
In particular, we could identify the ECTN distributions that contain a single motif, which were compatible with the topology of a simple connected infinite network.
Much work remains to be done in developing this framework, in particular finishing the study of Dirac distributions of ECTN and following the directions proposed in subsection \ref{subsec:tiling_th}.
Considering the results on the ECTN distribution, the following questions should be addressed in future work:
\begin{itemize}
    \item can the analytical formula obtained here account for the distribution observed for various facts known about centered motifs, like e.g. the histogram of the number of occurrences of NCTN and ECTN?
    This distribution has been first described in \cite{le2023modeling}, where it is referred to as ``aggregated weight for ETN''.
    Since this distribution is shared across many artificial and empirical face-to-face interaction data sets, explaining it would give us a deeper understanding of centered motifs.
    \item investigate how the expression of the NCTN and ECTN distributions change under time aggregation and local time shuffling of the temporal network.
    According to \cite{le2024flow}, this would give us knowledge about centered motifs that cannot be anticipated from their distributions sampled from the untouched data set.
    \item refine the principle of maximum entropy to take into account the non-stationarity of empirical networks.
    More precisely, the total number of active edges should be part of the constraints under which the entropy is maximized.
    \item Tuning an agent-based model to recover the NCTN distribution has been shown in \cite{le2023modeling} to synthesize more realistic artificial networks.
    By using the same tuning procedure as described in \cite{le2023modeling}, but recovering the ECTN distribution instead, further realism is expected.
    \item investigate whether a physical meaning can be associated to the hidden variables appearing in the hypothesis of conditional independence.
\end{itemize}

\newpage
\bibliography{main}

\clearpage
\newpage

\appendix

\setcounter{figure}{0}
\setcounter{equation}{0}
\setcounter{table}{0}
\setcounter{section}{0}
\renewcommand{\thefigure}{S\arabic{figure}}
\renewcommand{\thesection}{S\arabic{section}}
\renewcommand{\thetable}{S\arabic{table}}
\renewcommand{\theequation}{S\arabic{equation}}

\widetext

\section{\label{sec:AppendixA}recursion formula for the number of gluing matrices}

The goal of this section is to prove the following formula:
\begin{equation}\label{eq:gluing_mat}
B_{k,l}(n,m)=
\sum_{b\in\mathcal{B}_{i,l-j}(n,m)}
\sum_{b'\in\mathcal{B}_{k-i,j}(n,m)}
B_{i,j}\left[n[:i]-b_{.,+},m[:j]-b'_{+,.}\right]
B_{k-i,l-j}\left[n[i:]-b'_{.,+},m[j:]-b_{+,.}\right]
\end{equation}
See the sub-subsection \ref{subsubsec:gluing_nb} for the notations.
In particular, recall that $B_{k,l}(n,m)$ denotes the number of elements in $\mathcal{B}_{k,l}(n,m)$.
Let us consider $M$ such an element.
Then the constraints it must satisfy can be written as:
$$
\begin{cases}
    M_{.,+}\preceq n
    \\
    M_{+,.}\preceq m
\end{cases}
$$
where we have introduced the partial order $\preceq$ on vectors of integers:
$$
a\preceq b\iff\forall i,a_{i}\leq b_{i}
$$

\begin{figure}
    \includegraphics[width=0.5\columnwidth]{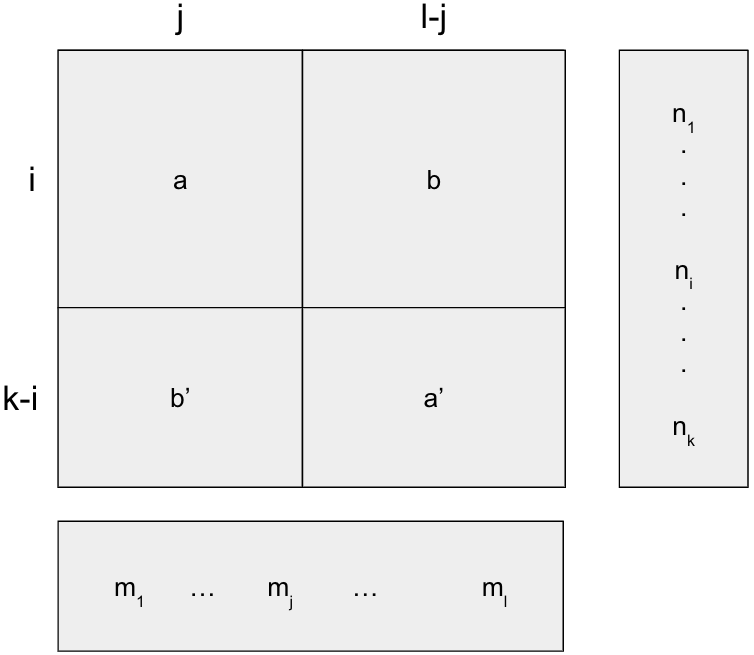}
    \caption{\textbf{Visual helper for the proof of eq \ref{eq:gluing_mat}}.
    The central block stands for the element $M$ discussed in the main text.
    The two vectors $n$ and $m$ have been placed next to $M$ to recall the constraints $M$ must satisfy:
    the sum over columns (left to right) is constrained by $n$ while the sum over rows (top to down) is constrained by $m$.
    As explained in the text, we divide $M$ into four sub-matrices $a,a',b,b'$ before rewriting the constraints on $M$ as constraints on $a,a',b$ and $b'$.
    }
    \label{fig:gluing_mat1}
\end{figure}

This notation allows us to considerably simplify the proof of eq \ref{eq:gluing_mat}.
Indeed, to count the number of possibilities for $M$, we divide $M$ into four sub-matrices $a,b,b',a'$ as depicted on Figure \ref{fig:gluing_mat1}.
On the figure, $a$ has $i$ rows and $j$ columns, etc.
From this figure, it is easy to rewrite the constraints satisfied by $M$ as:
$$
\begin{cases}
    a_{.,+}+b_{.,+}\preceq n[:i]
    \\
    a'_{.,+}+b'_{.,+}\preceq n[i:]
    \\
    a_{+,.}+b'_{+,.}\preceq m[:j]
    \\
    a'_{+,.}+b_{+,.}\preceq m[j:]
\end{cases}
$$

Thus, choosing an element $M$ in $\mathcal{B}_{k,l}(n,m)$ amounts to choose four matrices $a,b,a',b'$ with integer coefficients that satisfy the inequalities written above.
Using the equivalence $a\preceq b\iff a-b\preceq0$, we deduce that:
$$
\begin{cases}
    b_{.,+}\preceq n[:i]
    \\
    b_{+,.}\preceq m[j:]
\end{cases}
$$

Said otherwise, we have that $b\in\mathcal{B}_{i,l-j}(n[:i],m[j:])$.
Similarly, we have that $b'\in\mathcal{B}_{k-i,j}(n[i:],m[:j])$.
Then, for a given couple $(b,b')$, $a$ must satisfy:
$$
\begin{cases}
    a_{.,+}\preceq n[:i]-b_{.,+}
    \\
    a_{+,.}\preceq m[:j]-b'_{+,.}
\end{cases}
$$
Said otherwise, the number of possibilities for $a$ is equal to $B_{i,j}(n[:i]-b_{.,+},m[:j]-b'_{+,.})$.
Similarly, the number of possibilities for $a'$ is equal to $B_{k-i,l-j}(n[i:]-b'_{.,+},m[j:]-b_{+,.})$.

By summing over all the possible couples $(b,b')$, we obtain the desired formula \ref{eq:gluing_mat}.

\newpage
\section{\label{sec:AppendixC}sizes of the considered data sets}

\begin{table*}[b]
    \begin{tabular}{|c|c|c|c|}
        \hline
        data set name & number of nodes & duration & number of temporal edges \\
        \hline
        ``conf16'' & 138 & 3 635 & 153 371 \\
        \hline
        ``conf17'' & 274 & 7 250 & 229 536 \\
        \hline
        ``conf18'' & 164 & 3 519 & 96 362 \\
        \hline
        ``conf19'' & 172 & 7 313 & 132 949 \\
        \hline
        ``utah'' & 630 & 1 250 & 353 708 \\
        \hline
        ``french'' & 242 & 3 100 & 125 773 \\
        \hline
        ``highschool1'' & 126 & 5 609 & 28 560 \\
        \hline
        ``highschool2'' & 180 & 11 273 & 45 047 \\
        \hline
        ``highschool3'' & 327 & 7 375 & 188 508 \\
        \hline
        ``hospital'' & 75 & 9 453 & 32 424 \\
        \hline
        ``malawi'' & 86 & 43 438 & 102 293 \\
        \hline
        ``baboon'' & 13 & 40 846 & 63 095 \\
        \hline
        ``work1'' & 92 & 7 104 & 9 827 \\
        \hline
        ``work2'' & 217 & 18 488 & 78 249 \\
        \hline \hline
        ``brownD01'' & 1 000 & 4 000 & 635 266 \\
        \hline
        ``brownD001'' & 988 & 4 000 & 661 878 \\
        \hline
        ``ABP2pi'' & 201 & 4 000 & 17 698 \\
        \hline
        ``ABPpi4'' & 1 000 & 4 000 & 156 173 \\
        \hline
        ``Vicsek2pi'' & 500 & 4 000 & 38 137 \\
        \hline
        ``Vicsekpi4'' & 500 & 4 000 & 41 841 \\
        \hline \hline
        ``ADM9conf16'' & 138 & 3 635 & 187 774 \\
        \hline
        ``ADM18conf16'' & 138 & 3 635 & 86 793 \\
        \hline
        ``min\_ADM1'' & 138 & 3 635 & 154 628 \\
        \hline
        ``min\_ADM2'' & 138 & 3 635 & 151 078 \\
        \hline
        ``min\_EW1'' & 138 & 3 635 & 57 376 \\
        \hline
        ``min\_EW2'' & 138 & 3 635 & 193 203 \\
        \hline
        ``min\_EW3'' & 138 & 3 635 & 185 901 \\
        \hline
    \end{tabular}
    \caption{\label{tab:3}\textbf{Sizes of the 27 data sets considered in this paper.}
    }
\end{table*}

\newpage
\section{\label{sec:AppendixD}computation of the ECTN distribution under the hypothesis of independent nodes}
However, before considering the generic case, we will solve the case of depth 1.

\subsection{depth 1}
For an ECTN of depth 1, we will use a simpler notation than above because we have
$A_{n}(1)=\{`1'\}$ so a motif can be written as $[\vec{n}]$ instead of $(c,[\vec{n}])$.
As $A_{e}(1)=\{`1',`2',`3'\}$,
we denote by $P(n_{1},n_{2},n_{3})$ the probability that $i$ interacts with $j$, that $i$ interacts with $n_{1}$ exclusive partners, $j$ interacts with $n_{2}$ exclusive partners and $i$ and $j$ interact with $n_{3}$ common partners.

The two representations $\vec{n}^{(i)}$ and $\vec{n}^{(j)}$ are related by:
$$
\vec{n}^{(j)}=\left(n_{2}^{(i)},n_{1}^{(i)},n_{3}^{(i)}\right)
$$

We will also abbreviate $f(h_{i},h_{j})$ into $f_{i,j}$ for the sake of convenience.
Now, let us compute $P(\vec{n}|\vec{h})$.

We have:
\begin{multline}
    P(n_{1},n_{2},n_{3}|\vec{h})=f_{i,j}
    \sum_{I_{1},I_{2},I_{3},|I_{k}\cap I_{l}|=n_{k}\delta_{k,l}}
    \left(\prod_{l\in I_{1}}f_{i,l}(1-f_{j,l})\right)
    \left(\prod_{l\in I_{2}}f_{j,l}(1-f_{i,l})\right)\left(\prod_{l\in I_{3}}f_{i,l}f_{j,l}\right)
    \\
    \left(\prod_{l\notin I_{1}\cup I_{2}\cup I_{3}}(1-f_{i,l})(1-f_{j,l})\right)
\end{multline}

To make this expression more compact, let us introduce the following notations:
\begin{equation}
    \label{eq:1}
\begin{cases}
    \Lambda(N) = \prod_{l=1}^{N}(1-f_{i,l})(1-f_{j,l})
    \\
    f_{1}(h_{l})=\frac{f_{i,l}}{1-f_{i,l}}
    \\
    f_{2}(h_{l})=\frac{f_{j,l}}{1-f_{j,l}}
    \\
    f_{3}(h_{l})=\frac{f_{i,l}f_{j,l}}{(1-f_{i,l})(1-f_{j,l})}
    \\
    F_{k}(I)=\prod_{l\in I}f_{k}(h_{l})
\end{cases}
\end{equation}

Then we have:
$$
P((n_{k})_{1\leq k\leq3}|\vec{h})=
f_{i,j}
\Lambda(N-2)\sum_{I_{1},I_{2},I_{3},|I_{k}\cap I_{l}|=n_{k}\delta_{k,l}}
\prod_{k=1}^{3}F_{k}(I_{k})
$$

Now let us compute this formula in the case of a large number of nodes.
At first, let us consider the case of a single set $I$:
$$
S(f;n,N)=\sum_{I,I\subseteq\llbracket1,N\rrbracket\land|I|=n}F(I)
$$
where $F(I)=\prod_{l\in I}f(h_{l})$ and $f$ is an arbitrary real-valued function.

We have:
\begin{align*}
    S(f;n,N)
    &=
    \sum_{1\leq k_{1}<\cdots<k_{n}\leq N}\prod_{l=1}^{n}f(h_{k_{l}})
    \\
     &=
     \frac{1}{n!}\sum_{1\leq k_{1},\hdots,k_{n}\leq N|k_{l}\neq k_{l'},\forall l\neq l'}\prod_{l=1}^{n}f(h_{k_{l}})
     \\
     &= 
     \frac{1}{n!}\left[\left(\sum_{k=1}^{N}f(h_{k})\right)^{n}-\sum_{m=2}^{n}\sum_{k,k_{1},\hdots,k_{n-m}}
     f(h_{k})^{m}\prod_{l=1}^{n-m}f(h_{k_{l}})\right]
\end{align*}

Then we notice the following:
$$
\begin{cases}
    \sum_{k=1}^{N}f(h_{k})=
    N\left<f(h)\right>
    \\
    \sum_{k,k_{1},\hdots,k_{n-m}}f(h_{k})^{m}\prod_{l=1}^{n-m}f(h_{k_{l}})=\mathcal{O}\left(N^{n-m+1}\right)
\end{cases}
$$

We deduce that:
$$
S(f;n,N)=\frac{N^{n}}{n!}\left[\left<f(h)\right>^{n}+\mathcal{O}\left(\frac{1}{N}\right)\right]
$$

However, this result as such cannot be used for our computation of $P(n_{1},n_{2},n_{3}|h)$, since we need to consider the case of multiple sets $I_{k}$.
Let us compute:
\begin{align*}
    \hspace*{-1cm}S(f_{1},n_{1},f_{2},n_{2};N)
    &=
    \sum_{I_{1},I_{2}\subseteq\llbracket1,N\rrbracket,|I_{k}\cap I_{l}|=\delta_{k,l}n_{k}}F_{1}(I_{1})F_{2}(I_{2})
    \\
    &=
    S(f_{1},n_{1},N)S(f_{2},n_{2},N)
    -\sum_{n=1}^{\min(n_{1},n_{2})}
    \sum_{I_{1},I_{2},|I_{1}\cap I_{2}|=0,|I_{k}|=n_{k}-n}F_{1}(I_{1})F_{2}(I_{2})
    \\
    &~
    \sum_{J,|J|=n,|J\cap I_{k}|=0}F_{1}(J)F_{2}(J)
\end{align*}

We have:
$$
\begin{cases}
    \sum_{I_{1},I_{2}}F_{1}(I_{1})F_{2}(I_{2})\sum_{J}F_{1}(J)F_{2}(J)=\mathcal{O}\left(N^{n_{1}+n_{2}-n}\right)
    \\
    S(f_{1},n_{1},N)S(f_{2},n_{2},N)=N^{n_{1}+n_{2}}\left(c+\mathcal{O}\left(\frac{1}{N}\right)\right)
\end{cases}
$$

Whence the result we needed:
\begin{align*}
    S((f_{k},n_{k})_{1\leq k\leq d};N)
    &=
    \sum_{I_{1},\hdots,I_{d}\subseteq\llbracket1,N\rrbracket,|I_{k}\cap I_{l}|=\delta_{k,l}n_{k}}\prod_{k=1}^{d}F_{k}(I_{k})
    \\
    &=
    \prod_{k=1}^{d}S(f_{k},n_{k},N)+\cdots
    \\
    &=
    \frac{N^{\sum_{k=1}^{d}n_{k}}}{\prod_{k=1}^{d}n_{k}!}\prod_{k=1}^{d}\left<f_{k}(h)\right>^{n_{k}}+\cdots
\end{align*}

Applying this result to the functions $f_{1},f_{2},f_{3}$ defined in equation \ref{eq:1} yields:
\begin{align*}
    P((n_{k})_{1\leq k\leq3}|\vec{h})
    &=
    f_{i,j}\Lambda(N-2)\binom{N-2}{n_{1},n_{2},n_{3}}
    \prod_{k=1}^{3}\left<f_{k}(h)\right>^{n_{k}}
\end{align*}

To get rid of the binomial coefficient and the term $\Lambda$, which are model-specific, we introduce the distribution for the number of satellites:
\begin{align*}
    \Omega(n|\vec{h})
    &=
    \sum_{I\subseteq\llbracket1,N-2\rrbracket,|I|=n}\prod_{l\in I}\left[1-(1-f_{i,l})(1-f_{j,l})\right]
    \prod_{l\notin I}(1-f_{i,l})(1-f_{j,l})
    \\
    &\simeq
    \Lambda(N-2)\binom{N-2}{n}\left<f_{\Omega}\right>^{n}
\end{align*}
where
$$
f_{\Omega}(h_{l})=\frac{1}{(1-f_{i,l})(1-f_{j,l})}-1
$$

We deduce the following formula:
\begin{align}
    P(n_{1},n_{2},n_{3}|h)
    &\simeq
    f_{i,j}\Omega(n|h)n!
    \prod_{k=1}^{3}\frac{p_{k}^{n_{k}}}{n_{k}!}
\end{align}
where $n=\sum_{k=1}^{3}n_{k}$ and we have introduced
$$
p_{k}=\frac{\left<f_{k}(h)\right>}{\left<f_{\Omega}(h)\right>}
$$
Note that we have $\sum_{k=1}^{3}p_{k}=1$ and that the $p_{k}$ depend on $h_{1},\hdots,h_{N}$.
However, in the limit $N\xrightarrow{}\infty$, they depend only on $h_{i}$ and $h_{j}$, since the sampling average $<>$ identifies with the expected value $\mathop{\mathbb{E}}()$.
For the same reason, we have
$$
\Omega(n|\vec{h})\underset{N\xrightarrow{}\infty}{=}\Omega(n|h_{i},h_{j})
$$

This leads to
\begin{align*}
    P(n_{1},n_{2},n_{3})=
    \sum_{h_{i},h_{j}}\rho(h_{i})\rho(h_{j})
    P(n_{1},n_{2},n_{3}|h_{i},h_{j})
\end{align*}

We deduce the complete expression for the probability of a motif $[\vec{n}]$ of depth 1.\\
If $n_{1}\neq n_{2}$:
\begin{align*}
    P([\vec{n}])
    &=
    \binom{n}{n_{1},n_{2}}
    \sum_{h,h'}\rho(h)\rho(h')f(h,h')
    \Omega(n|h,h')p_{3}(h,h')^{n_{3}}
    \left[
    p_{1}(h,h')^{n_{1}}
    p_{2}(h,h')^{n_{2}}+
    p_{1}(h,h')^{n_{2}}p_{2}(h,h')^{n_{1}}
    \right]
\end{align*}
If $n_{1}=n_{2}$:
\begin{align*}
    P([\vec{n}])
    &=
    \binom{n}{n_{1},n_{1}}
    \sum_{h,h'}\rho(h)\rho(h')f(h,h')
    \Omega(n|h,h')
    p_{3}(h,h')^{n_{3}}
    \left[
    p_{1}(h,h')p_{2}(h,h')
    \right]^{n_{1}}
\end{align*}

Excluding the ECTN with an empty profile for the central edge leads to:
$$
P^{*}([\vec{n}])=
\frac{P([\vec{n}])}{\sum_{h,h'}\rho(h)\rho(h')f(h,h')}
$$

\subsection{generic depth}
Now we want to compute the ECTN probability under the hypothesis of independent nodes for a generic depth $d$.
As before, let us consider an edge $i,j$ that will constitute the central edge.
We have:
\begin{align*}
    P(c,\vec{n}|h_{i},h_{j})
    &=
    f_{c}(c)
    \sum_{(I_{a})_{a\in A_{e}(d)},|I_{a}\cap I_{a'}|=n_{a}\delta_{a,a'}}
    \prod_{a\in A_{e}(d)}F_{a}(I_{a})
    \prod_{l\notin\cup_{a\in A_{e}(d)}I_{a}}f_{\emptyset}(h_{l})
\end{align*}
where $f_{c}(c)$ denotes the probability that the central edge has $c$ as profile and $f_{\emptyset}(h)$ denotes the probability that a node with state $h$ has no interactions with either $i$ or $j$.
$F_{a}(I)$ denotes the probability that all the nodes in $I$ have the same profile $a$.
Note that $F_{a}$, $f_{\emptyset}$ and $f_{c}$ are all conditioned to $h_{i}$ and $h_{j}$.
Introducing $\Lambda(N)=\prod_{l=1}^{N}f_{\emptyset}(h_{l})$ we get (cf above)
\begin{align*}
    \hspace*{-0.5cm}P(c,\vec{n}|h_{i},h_{j})
    &\simeq
    f_{c}(c)\Lambda(N-2)\binom{N-2}{(n_{a})_{a\in A_{e}(d)}}
    \prod_{a\in A}\left<\tilde{f}_{a}(h)\right>^{n_{a}}
\end{align*}
where $\tilde{f}_{a}=\frac{f_{a}}{f_{\emptyset}}$

On the other hand we have:
\begin{align*}
    \Omega(n|h_{i},h_{j})
    &\simeq
    \Lambda(N-2)\binom{N-2}{n}\left<f_{\Omega}(h)\right>^{n}
\end{align*}
where $f_{\Omega}=\frac{1}{f_{\emptyset}}-1$

Hence
\begin{align*}
    P(c,\vec{n}|h_{i},h_{j})
    &\simeq
    f_{c}(c|h_{i},h_{j})\Omega(n|h_{i},h_{j})n!
    \prod_{a\in A}\frac{p_{a}(h_{i},h_{j})^{n_{a}}}{n_{a}!}
\end{align*}
where $n=\sum_{a\in A_{e}(d)}n_{a}$ and we have introduced
$$
p_{a}(h_{i},h_{j})=\frac{\left<\tilde{f}_{a}(h)\right>}{\left<f_{\Omega}(h)\right>}
$$

The complete expression for the ECTN probability writes:
\begin{align*}
    P(c,[\vec{n}])
    &=
    \binom{n}{\vec{n}}\sum_{h,h'}\rho(h)\rho(h')f_{c}(c|h,h')
    \Omega(n|h,h')
    \sum_{\vec{n}\in[\vec{n}]}
    \prod_{a\in A_{e}(d)}
    p_{a}(h,h')^{n_{a}}
\end{align*}
Excluding the ECTN with an empty profile for the central edge leads to:
$$
P^{*}(c,[\vec{n}])=
\frac{P(c,[\vec{n}])}{1-\sum_{h,h'}\rho(h)\rho(h')f_{c}(\emptyset|h,h')}
$$

\newpage
\section{proving the non-universality of the hypothesis of conditional independence\label{sec:cond_ind_test}}
Recall that any distribution satisfying the hypothesis of conditional independence can take the form \ref{eq:10}, that we recall here:
\begin{equation}\label{eq:100}
\begin{cases}
    g(\vec{n})
    =
    \binom{n}{\vec{n}}
    \sum_{e}
    \Omega_{e}(n)\prod_{k=1}^{D}p_{e,k}^{n_{k}}
    \\
    \Omega_{e}(n),p_{e,k}
    \geq0,\forall e,n,k
    \\
    \sum_{k}p_{e,k}
    =
    1,\forall e
\end{cases}
\end{equation}
where $D\geq2$, $\vec{n}=(n_{1},\hdots,n_{D})$ and $n=\sum_{k}n_{k}$.

More generally, we consider $g$ to be an integrable function defined on tuples of integers, where $g$ integrable means that
$$\sum_{n_{1}=0}^{\infty}\cdots\sum_{n_{d}=0}^{\infty}g(\vec{n})<\infty$$
The integrability allows $g$ to be interpreted as a distribution, since it is positive and can be normalized.

We have:
\begin{align*}
\hspace{-3em}
    \sum_{\vec{n}}g(\vec{n})
    &=
    \sum_{e}\sum_{n=0}^{\infty}\Omega_{e}(n)
    \sum_{n_{1},\hdots,n_{d-1}=0}^{\infty}
    \binom{n}{n_{1},\hdots,n_{d-1}}
    \prod_{k=1}^{D}p_{e,k}^{n_{k}}
    \\
    &=
    \sum_{e}\sum_{n=0}^{\infty}\Omega_{e}(n)\left(\sum_{k=1}^{D}p_{e,k}\right)^{n}
    \\
    &=
    \sum_{e}\sum_{n=0}^{\infty}\Omega_{e}(n)
\end{align*}

Let us define
$$
\lambda_{e}=\sup_{n\in\mathbb{N}}\Omega_{e}(n)
$$

As $\forall e,n,\Omega_{e}(n)\geq0$ we deduce that
$$
\sum_{e}\lambda_{e}<\infty
$$

How $g(\vec{n})$ varies as one of the components of $\vec{n}$ grows while the other are held constants?

Since all the terms in the sum of eq \ref{eq:100} are positive, $g(\vec{n})$ is invariant under any permutation of the hidden states.
Thus we can assume that the sequence $(p_{e,1})_{e}$ is increasing monotonously.
Since it is bounded, it must converge so let us denote its limit by $p_{1}$.

We will consider two cases for $p_{1}$:
\begin{enumerate}
    \item $p_{1}<1$
    \item $p_{1}=1$
\end{enumerate}

In the general case, we can write
\begin{align*}
    g(\vec{n})
    &<
    \frac{n!}{n_{k}!}\sum_{e}\lambda_{e}p_{e,k}^{n_{k}}\prod_{l\neq k}\frac{p_{e,l}^{n_{l}}}{n_{l}!}
    \\
    &\leq
    \frac{n!}{n_{k}!}\sum_{e}\lambda_{e}p_{e,k}^{n_{k}}
\end{align*}
where the term on the right side is finite because every term is positive and their sum is lower that $\sum_{e}\lambda_{e}$, which is itself finite.

Now, let us consider $n_{k}\xrightarrow{}\infty$ while the other components of $\vec{n}$ are fixed.
We have:
\begin{align*}
    \frac{(n_{k}+m)!}{n_{k}!}
    &=
    \prod_{i=1}^{m}\left(n_{k}+i\right)
    \underset{n_{k}\xrightarrow{}\infty}{\sim}
    n_{k}^{m}
\end{align*}

So we have some constant $C$ such that
\begin{align*}
    g(\vec{n})<Cn_{k}^{m}\sum_{e}\lambda_{e}p_{e,k}^{n_{k}}
\end{align*}
where $m=\sum_{l\neq k}n_{l}$

Now, if $p_{1}<1$, we deduce that
\begin{align*}
    g(\vec{n})
    &<
    Cn_{1}^{m}p_{1}^{n_{1}}\sum_{e}\lambda_{e}
    \\
    &<C'n_{1}^{m}p_{1}^{n_{1}}
\end{align*}
In particular, when $n_{1}\xrightarrow{}\infty$ while the $n_{k\neq1}$ are held constant, we have that:
\begin{equation}\label{eq:11}
    \forall m\in\mathbb{N},g(\vec{n})n_{1}^{m}\xrightarrow{}0
\end{equation}

In the case $p_{1}=1$, we have that $\forall k\neq1,p_{e,k}\xrightarrow{}0$ since
$$
0\leq p_{e,k}\leq\sum_{l\neq1}p_{e,l}=1-p_{e,1}\xrightarrow{}0
$$
Then, we have the following:
$$
\sup_{e,p_{e,2}<1}p_{e,2}<1
$$
Moreover, for any $e$ such that $p_{e,2}=1$ we have e.g. $p_{e,1}=0$ so taking $n_{1}=1$ is enough to remove the terms $e$ such that $p_{e,2}=1$ from the sum \ref{eq:100}.
Thus, if we consider $n_{2}\xrightarrow{}\infty$ while the $n_{k\neq2}$ are held constant with $n_{1}\neq0$, we have:
\begin{align*}
    g(\vec{n})
    &=
    \binom{n}{\vec{n}}
    \sum_{e,p_{e,2}<1}\Omega_{e}(n)\prod_{k}p_{e,k}^{n_{k}}
    <C'n_{2}^{m}p_{2}^{n_{2}}
\end{align*}
where $m=\sum_{k\neq2}n_{k}$ and we introduced
$$
p_{2}
=
\sup_{e,p_{e,2}<1}p_{e,2}
$$
Since $p_{2}<1$, we have that:
\begin{equation}\label{eq:12}
    \forall m\in\mathbb{N},g(\vec{n})n_{2}^{m}\xrightarrow{}0
\end{equation}
when $n_{2}\xrightarrow{}\infty$ while the $n_{k\neq2}$ are held constant, and at least one of the $n_{k\neq2}$ is not zero.

From equations \ref{eq:11} and \ref{eq:12}, we see that not any positive integrable function $g$ can be written under the form \ref{eq:100}, even when the sum over the hidden states $e$ is infinite countable.

For example, let us consider $g$ so that:
$$
g(\vec{n})=f(n)f_{1}(n_{1})
$$
with $f$ and $f_{1}$ decreasing as powers of $n$.
For example, we can take
$$
\begin{cases}
    f(n)=\frac{1}{n^{D+1}+1}
    \\
    f_{1}(n_{1})=\frac{1}{n_{1}+1}
\end{cases}
$$

Then, either $p_{1}<1$ or $p_{1}=1$.
If $p_{1}<1$, eq \ref{eq:11} leads to:
$$
f_{1}(n_{1})n_{1}\underset{n_{1}\xrightarrow{}\infty}{\xrightarrow{}}0
$$
but $f_{1}(n_{1})n_{1}\xrightarrow{}1$.
If $p_{1}=1$, eq \ref{eq:12} leads to $f_{1}(n_{1})=0,\forall n_{1}>0$, which is not the case.

Thus, at least in the case of a countable number of hidden variables $e$, we have the certainty that not any distribution can be approximated by the generic form \ref{eq:100}.

\end{document}